\documentclass[times,5p]{elsarticle}

\usepackage{layouts}
\usepackage{microtype}

\usepackage{booktabs}
\usepackage{makecell}

\usepackage[dvipsnames]{xcolor}
\usepackage{amsmath,amssymb,bm}
\usepackage{xspace}
\usepackage{standalone}

\usepackage{listings}
\usepackage[pdfencoding=unicode,linkcolor=blue,citecolor=blue,urlcolor=blue]{hyperref}
\usepackage[subrefformat=parens,labelfont=bf,font={stretch=1.15}]{subcaption}

\usepackage[capitalise]{cleveref}
\crefname{appendix}{\unskip}{\unskip} 

\usepackage{tikz}
\usetikzlibrary{positioning,patterns,patterns.meta,shapes.geometric,arrows.meta,backgrounds,intersections,decorations.pathmorphing,calc,through}
\usepackage{siunitx}[=v2]

\journal{Astroparticle Physics}

\usepackage{xspace}

\newcommand{\cor}[1]{CORSIKA~{#1}\xspace}

\newcommand{\Offline}{\mbox{$\overline{\textrm{Off}}$\hspace{.05em}%
    \raisebox{.4ex}{$\underline{\textrm{line}}$}}\xspace}

\newcommand{\lambdaInt}{\lambda_{\mathrm{int}}}

\newcommand{\hScale}{h_{\mathrm{sc}}}

\newcommand{\Xmax}{\ensuremath{X_{\mathrm{max}}}\xspace}

\newcommand{\Eth}{E_{\mathrm{th}}}

\newcommand{\wmax}{w_{\mathrm{max}}}

\DeclareSIUnit\yr{yr}

\begin{document}
\begin{frontmatter}
\title{\cor8: A General Framework for Particle Cascade Simulations}

\author[1,2]{Jean-Marco~Alameddine}
\author[1,2]{Johannes~Albrecht}
\author[3,4]{Antonio Augusto~Alves Jr.}
\author[5,6]{Juan~Ammerman-Yebra}
\author[7]{Luisa~Arrabito}
\author[1,2]{Dominik~Baack}
\author[8,9]{Alan~Coleman}
\author[10]{Cosmin~Deaconu}
\author[1,2]{Hans~Dembinski}
\author[1,2]{Dominik~Elsässer}
\author[3]{Ralph~Engel}
\author[7]{Alice~Faure}
\author[3]{Alfredo~Ferrari}
\author[11]{Chloé~Gaudu}
\author[8,1,2]{Christian~Glaser}
\author[3]{Marvin~Gottowik}
\author[3]{Dieter~Heck}
\author[3,12]{Tim~Huege}
\author[11]{Karl-Heinz~Kampert}
\author[3]{Nikolaos~Karastathis}
\author[13]{Jeffrey~Lazar}
\author[14]{Lukas~Nellen}
\author[15,16]{David~Parello}
\author[3]{Tanguy~Pierog}
\author[17]{Remy~Prechelt}
\author[18,19,20]{Radek~Privara}
\author[21,9]{Maximilian~Reininghaus}
\author[1,2]{Wolfgang~Rhode}
\author[1,22,6]{Felix~Riehn}
\author[1,2]{Maximilian~Sackel}
\author[3]{Pranav~Sampathkumar}
\author[11]{Alexander~Sandrock}
\author[3]{André~Schmidt}
\author[1,2]{Jan~Soedingrekso}
\author[3]{Ralf~Ulrich}
\author[10]{Philipp~Windischhofer}
\author[11]{Baobiao~Yue}

\affiliation[1]{organization={Technische Universität Dortmund (TU), Department of Physics}, city={Dortmund}, country={Germany}}
\affiliation[2]{organization={Lamarr Institute for Machine Learning and Artificial Intelligence}, city={Dortmund}, country={Germany}}
\affiliation[3]{organization={Karlsruhe Institute of Technology (KIT), Institute for Astroparticle Physics (IAP)}, city={Karlsruhe}, country={Germany}}
\affiliation[4]{organization={University of Cincinnati, Cincinnati}, city={OH}, country={United States}}
\affiliation[5]{organization={IMAPP, Radboud University Nijmegen}, city={Nijmegen}, country={The Netherlands}}
\affiliation[6]{organization={Universidade de Santiago de Compostela, Instituto Galego de Física de Altas Enerxías (IGFAE)}, city={Santiago de Compostela}, country={Spain}}
\affiliation[7]{organization={Laboratoire Univers \& Particules de Montpellier, CNRS \& Université de Montpellier (UMR-5299)}, city={34095 Montpellier}, country={France}}
\affiliation[8]{organization={Uppsala University, Department of Physics and Astronomy}, city={Uppsala}, country={Sweden}}
\affiliation[9]{Independent researcher}
\affiliation[10]{organization={Department of Physics, Enrico Fermi Institute, Kavli Institute for Cosmological Physics, University of Chicago, Chicago}, city={IL 60637}, country={USA}}
\affiliation[11]{organization={Bergische Universität Wuppertal, Department of Physics}, city={Wuppertal}, country={Germany}}
\affiliation[12]{organization={Vrije Universiteit Brussel, Astrophysical Institute}, city={Brussels}, country={Belgium}}
\affiliation[13]{organization={UCLouvain, Centre for Cosmology, Particle Physics and Phenomenology, CP3, Chemin du Cyclotron 2}, city={1348 Louvain-la-Neuve}, country={Belgium}}
\affiliation[14]{organization={Universidad Nacional Autónoma de México (UNAM), Instituto de Ciencias Nucleares}, city={México}, country={México}}
\affiliation[15]{organization={DALI, Univ Perpignan}, city={Perpignan}, country={France}}
\affiliation[16]{organization={LIRMM Univ Montpellier, CNRS}, city={Montpellier}, country={France}}
\affiliation[17]{organization={University of Hawai'i at Manoa, Department of Physics and Astronomy}, city={Honolulu}, country={USA}}
\affiliation[18]{organization={European Southern Observatory (ESO)}, city={Garching}, country={Germany}}
\affiliation[19]{organization={Palacký University in Olomouc, Faculty of Science, Joint Laboratory of Optics}, city={Olomouc}, country={Czech Republic}}
\affiliation[20]{organization={Institute of Physics of the Academy of Sciences of the Czech Republic, Joint Laboratory of Optics}, city={Olomouc}, country={Czech Republic}}
\affiliation[21]{organization={Karlsruhe Institute of Technology (KIT), Institut für Experimentelle Teilchenphysik (ETP)}, city={Karlsruhe}, country={Germany}}
\affiliation[22]{organization={Laboratório de Instrumentação e Física Experimental de Partículas (LIP)}, city={Lisboa}, country={Portugal}}
\begin{abstract}
The simulation of extensive air showers and particle cascades in general is a cornerstone of modern astroparticle physics. For more than two decades, CORSIKA, currently in version 7, has been one of the most widely used tools for this purpose. However, its architecture reflects design constraints of an earlier computing era, as well as increasingly limiting extensibility, maintainability, and adaptability to modern experimental requirements. \cor8 is a complete redesign of the original CORSIKA code, implemented in modern C++ and based on contemporary software engineering principles. It introduces a modular and extensible simulation framework with explicit handling of units, flexible geometry, and environment descriptions. In this paper, we present the design philosophy and core architecture of \cor8, describe the implementation of electromagnetic and hadronic shower physics, and validate air shower simulations against \cor7. The results demonstrate good agreement at the few-percent level for key observables, confirming the physics fidelity of \cor8. We also showcase new use cases that were beyond the capabilities of version 7, such as the simulation of cross-media showers and particle cascades in ice, including radio-signal propagation.
\end{abstract}
\end{frontmatter}


\section{Introduction}

The simulation of extensive air showers (EAS), as well as particle cascades in other media, is a cornerstone of modern astroparticle physics. Ground-based experiments rely on detailed Monte Carlo simulations both when air showers constitute the primary signal, as in cosmic-ray and gamma-ray observatories, and when they form an irreducible background, as in neutrino and multi-messenger experiments. Accurate air-shower simulations are indispensable for detector design, event reconstruction, energy and mass composition measurements, and for the interpretation and comparison of results across experiments.

For more than two decades, several mature Monte Carlo simulation programs have been used to simulate extensive air showers in the atmosphere. Among the most widely used are CORSIKA~\cite{Heck:1998vt} and AIRES~\cite{aires_reference}. These codes have been adopted by a broad range of cosmic-ray, gamma-ray, and neutrino detection experiments and have played a central role in the interpretation of air-shower measurements.

The current version of CORSIKA, commonly referred to as \cor7, is still being actively maintained and continues to deliver excellent physics performance. However, its software architecture reflects design decisions rooted in the constraints of
FORTRAN~77 and subsequent extensions. Fortran \emph{common blocks}, roughly equivalent to global variables in other programming languages and used ubiquitously in \cor7, have been declared obsolescent in the Fortran~2018 standard~\cite{Reid:2018}. Even though only minor adaptations were needed during the 30-year history of \cor7 to ensure its buildability on ever-evolving platforms, the obsolescence of legacy FORTRAN may render the code unusable in the (admittedly distant) future.

At the same time, the requirements placed on air-shower simulations by current and future experiments are evolving rapidly. Modern observatories demand increased flexibility in the choice and combination of physics models, support for heterogeneous detector geometries and media, improved interfaces to analysis frameworks, and efficient exploitation of contemporary computing architectures. While the highly optimized and monolithic structure of \cor7 delivers excellent performance for the well-established use case of particle cascades in air, it significantly complicates the implementation, validation, and long-term maintenance of new features. Furthermore, the pool of developers with deep expertise in large FORTRAN codebases is steadily shrinking, posing an additional risk to the sustainability of the software ecosystem.

Motivated by these considerations, \cor8 was developed as a complete redesign of the CORSIKA core, based on modern C++ and contemporary software engineering principles~\cite{Engel:2018akg}. Rather than a line-by-line translation, \cor8 provides a modular, extensible, and transparent simulation framework. It preserves the well-tested physics concepts of its predecessor, enables new use cases that are challenging to address with \cor7, and ensures long-term maintainability.
The design emphasizes a modular structure in which each component -- such as physics models, geometry definitions, particle transport, and output handling -- is implemented independently through well-defined interfaces. This allows components to be developed, tested, or replaced without affecting the rest of the framework, and enables flexible combinations of physics models, geometries, and output formats.

\cor8 is being developed as a fully open-source project \cite{corsika8_gitlab} and is intended as a long-term community effort. Contributions in the form of code, physics models, validation studies, documentation, and user feedback are greatly appreciated and actively encouraged. At the time of writing, \cor8 is considered ``physics complete'' for the simulation of extensive air showers. Stable, versioned releases are being published on Zenodo~\cite{corsika8_zenodo_all} in the form of Apptainer containers~\cite{Kurtzer:2017}, enabling straightforward out-of-the-box usage on a wide range of computing platforms.

In this article, we present an overview of the motivation, design philosophy, and current status of \cor8. We describe the architectural concepts and core components of the framework, summarize the implemented physics and technical features, and present validation studies comparing \cor8 air-shower simulations with \cor7 (version 7.75). We also illustrate the capabilities of \cor8 beyond traditional air-shower applications by presenting selected example results. The results presented in the following sections were obtained using the ``ICRC 2025 release'' version of \cor8~\cite{corsika8_zenodo_icrc25}.

\section{Architecture}
In a very simplified manner, \cor8 can be described as performing mainly two tasks: propagation of individual particles under the influence of electromagnetic forces and energy losses, and the handling of stochastic events, i.e., decays and interactions, that involve the generation of new particles.

The first task involves solving the respective equations of motion, typically via numerical methods, taking into account constraints that limit the range of propagation such as geometrical and energy cuts. Moreover, the range of propagation is limited by the occurrence of stochastic events. Care must be taken to ensure that these stochastic ranges of propagation are correctly sampled from their respective distributions,
which often do not have a closed-form analytic expression (e.g., a particle with an energy-dependent interaction cross-section propagating in a medium with inhomogeneous density while suffering from continuous energy losses).

The physically accurate event generation is itself a huge task on its own. It
requires careful modeling of the relevant physics and is therefore outsourced to specialized codes, such as hadronic or electromagnetic interaction models that provide on the one hand
cross sections of the respective processes and on the other hand the machinery to sample such an exclusive event with the full kinematic state of the produced secondary particles.

\begin{figure}
    \centering
    \includegraphics[width=0.99\columnwidth]{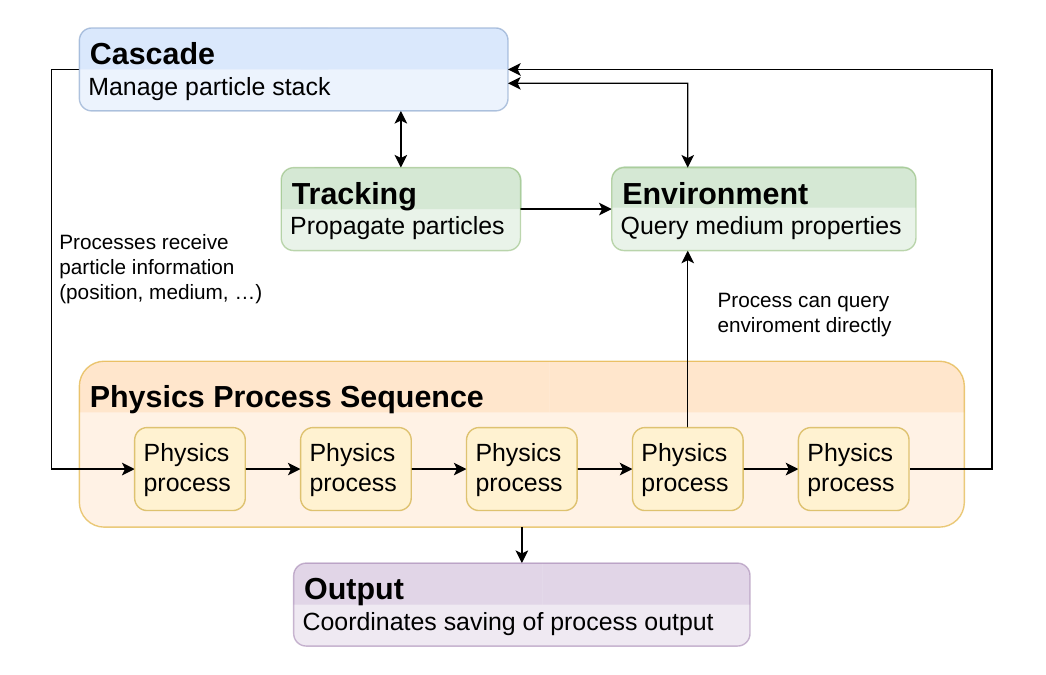}
    \caption{General structure of the \cor8 code. The Cascade manages the particle stack and main loop, invoking tracking and a configurable sequence of physics processes. Processes may query environmental properties directly, and output modules collect and store simulation results.}
    \label{fig:flowchart}
\end{figure}

However, these two aspects alone would be pointless to implement without proper \emph{bookkeeping} facilities that gather and finally store the data users are interested in, such as particle distributions on the ground and longitudinal profiles. Figure~\ref{fig:flowchart} shows how these components are arranged and interact within the overall structure of \cor8. In the following, we will explain the components in \cor8 in a ``bottom-up'' manner. Further details are given in Ref.~\cite{Reininghaus:2022qyl}.

\subsection{Fundamentals}
First, we discuss several fundamental cornerstones used in the design of \cor8.

\subsubsection{Unit system}\label{sec:units}
Incorrect usage of physical units of measure is regarded one of the three
most common errors in scientific computing~\cite{ISO-Fortran-Units}. 

Especially in a collaborative environment, such as \cor8,
a strategy to avoid these kinds of error is a key asset, also because their presence may render the simulation
wrong in an inconspicuous manner. One possible solution is to annotate data types with units. This solution was proposed more
than 40~years ago~\cite{Gehani:1977}. However, hardly any programming language today provides built-in support, so custom libraries are usually required~\cite{McKeever:2020}. In C++ the technique of template metaprogramming can be used to implement a compile-time dimensional analysis~\cite{Umrigar:1994}. In \cor8, we make use of the \textit{PhysUnits C++11} library~\cite{PhysUnits}, which employs this technique, with some custom modifications.

The basic idea is to introduce individual data types for each dimensionful quantity occurring in the code (length, time, density, etc.) and
enforce their usage instead of plain floating-point numbers. The multiplication and division of dimensionful quantities yields quantities
with different dimensions, so that an arbitrary number of such data types must be defined. In C++ this can readily be achieved
with template metaprogramming:

A templated type \lstinline{dimensions<N1, N2, N3, N4, N5, N6, N7, N8>} is introduced to keep track
of the dimensions of a quantity. The first seven integers $N_1,\ldots,N_7$ represent the exponents of the basic physical dimensions
\emph{length}, \emph{mass}, \emph{time}, \emph{electric current}, \emph{absolute temperature}, \emph{amount of substance}, and \emph{luminous intensity}.
The final integer represents the exponent of a custom \emph{HEP energy dimension} whose purpose will be explained below.

The actual type for quantities, \lstinline{quantity<Dim, Rep=double>}, is likewise templated. The first template argument is a \lstinline{dimensions} type,
the second argument selects the underlying floating point type that is used to store the numerical value of the quantity, by default \lstinline{double},
in its base unit. For better readability, type aliases like \lstinline{LengthType}, \lstinline{TimeType} or \lstinline{GrammageType} are defined so that one usually does not
need to handle the dimension indices explicitly.

The laws of quantity calculus (see, e.g., Ref.\ \cite{Raposo:2018}) are encoded in C++ code. For instance, only quantities of the same dimension
can be added and multiplying a \lstinline{quantity<dimensions<N1, N2, N3, ...>>} with a \lstinline{quantity<dimensions<M1, M2, M3, ...>>} returns a
\lstinline{quantity<dimensions<N1+M1, N2+M2, N3+M3, ...>>}.
This way, dimensional analysis of all calculations involving quantities
is conducted during compilation and any violation of the laws of quantity calculus results in a compiler error.

Besides the dimensional analysis, also the conversion of units to common base units is performed. A number of predefined constants
are provided, as well as \emph{user-defined literals} for convenience to initialize a quantity. For example, writing \lstinline{height = 1.450_km}
(where \lstinline{height} is a \lstinline{quantity} for length) or \lstinline{B = 0.48 * gauss + 5_uT} (where \lstinline{B} is a \lstinline{quantity}
for magnetic flux density) converts the quantity from the value in the stated unit to the base unit.

The strict dimensional analysis based on the SI becomes inconvenient when switching to the natural units
of particle physics, where the convention $c = \hbar = 1$ is employed, which reduces the number of dimensions by two. The unit system
as described so far prevents the use of intentionally ``sloppy'' expressions, such as $E^2 = p^2 + m^2$, in code. This is remedied by the introduction
of the spurious eighth HEP energy dimension, which we treat as independent. Quantities of this type (\lstinline{HEPEnergyType}) are intended for particle masses, energies
and momenta, expressed in multiples of electronvolts. Only a few situations in \cor8 exist where these microscopic units come into contact with
the macroscopic SI units. For these cases, conversion functions are provided to convert back and forth between SI and natural units when possible,
which mainly relies on the identity $\hbar c = \SI{197.327}{\MeV\femto\metre}$, which relates energy to length. To convert a quantity $q$ whose
dimensions are $\mathrm{length}^l \times \mathrm{time}^t \times \mathrm{mass}^m$ from SI to natural units, the conversion reads
\begin{equation}
\frac{q}{\si{eV}^{m - t - l}} = \left(\frac{\hbar c}{\si{\eV\metre}}\right)^{m - t - l} \cdot
    \left(\frac{\hbar}{\si{\kg\metre\squared\per\second}}\right)^{-m} \cdot
    \left(\frac{c}{\si{\metre\per\second}}\right)^{m+t} \times 
    \frac{q}{\si{\metre}^l \, \si{\second}^t \, \si{\kg}^m},
\end{equation}
so that the final dimensions are $\mathrm{energy}^{m - t - l}$. The inverse conversion of a quantity in natural units with dimension $\mathrm{energy}^e$ reads
\begin{equation}
\frac{q}{\si{\metre}^l \, \si{\second}^t \, \si{\kg}^m} = \left(\frac{\hbar c}{\si{\eV\metre}}\right)^{-e} \cdot
    \left(\frac{\hbar}{\si{\kg\metre\squared\per\second}}\right)^{m} \cdot
    \left(\frac{c}{\si{\metre\per\second}}\right)^{t-m} \times 
    \frac{q}{\si{eV}^e},
\end{equation}
with the constraint $m - l - t = e$.

\subsubsection{Geometry}\label{sec:geometry}
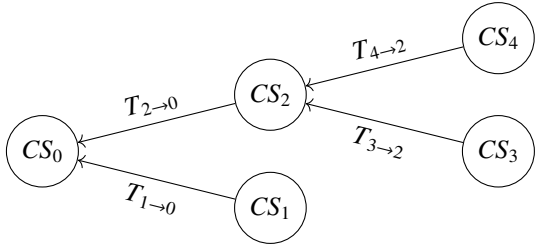
\begin{figure}
\centering
\begin{tikzpicture}[
	csnode/.style = {draw, shape=circle},
	edge from parent/.style = {draw, <-},
    level distance = 3cm,
    sloped
]
  \node [csnode] {$\mathit{CS}_0$} [grow=right]
    child { node [csnode] {$\mathit{CS}_1$} edge from parent node [below] {$T_{1 \rightarrow 0}$} }
    child { node [csnode] {$\mathit{CS}_2$} 
            child { node [csnode] {$\mathit{CS}_3$} edge from parent node [below] {$T_{3 \rightarrow 2}$} }
            child { node [csnode] {$\mathit{CS}_4$} edge from parent node [above] {$T_{4 \rightarrow 2}$} }
            edge from parent node [above] {$T_{2 \rightarrow 0}$} };
\end{tikzpicture}
\caption{Example coordinate system tree. $\mathit{CS}_0$ is the root coordinate system.}
\label{fig:cstree}
\end{figure}

The second cornerstone of \cor8 is its geometry classes that deal with points, vectors, and coordinate systems. The design borrows ideas from the
\Offline software framework of the Pierre Auger Collaboration~\cite{Argiro:2007qg}. Points and vectors are modeled not just as 3-dim.\ tuples of
their coordinates/components, but are always defined with respect to a specific coordinate system (CS), of which there can be multiple. Having several
CSs is useful for instance when interfacing event generators, which usually follow the convention to align the momenta of the projectile
particle along the $z$-axis. One unique root CS is predefined and new CSs are defined in terms of a reference CS and a transformation matrix that relates
the new CS with its reference. Supported transformations are elements of the special Euclidean group $\mathrm{SE}(3)$, i.e., translations and rotations,
represented by $4 \times 4$ transformation matrices.
These come into play when operations involve two vectors or points defined in different CSs. In this case, the components/coordinates of one of the objects are
temporarily transformed into the CS of the other object. Since the CSs form a tree structure, as illustrated in Fig.~\ref{fig:cstree}, this is achieved by multiplying
the transformation matrices along the shortest path connecting the two CSs. For example, the transformation required to go from $\mathit{CS}_3$ to $\mathit{CS}_1$
is $T_{3 \rightarrow 1} = T_{3 \rightarrow 2} T_{2 \rightarrow 0} T_{1 \rightarrow 0}^{-1}$. These transformations happen completely transparently to
the developer and the explicit usage of CS is barely necessary. Most often, expressions can be very close to the symbolic notation used in equations
as well. Explicit usage of coordinates
is required mostly during the initialization phase of a simulation, when the geometry is set up. For the actual linear algebra computations, the geometry system relies on the Eigen3 library~\cite{Eigen}, which is highly optimized.

We clearly distinguish between points and vectors as they appear in \emph{affine space} (see e.g.\ Ref.~\cite{Shafarevich:2013}): Points
are subject to rotations and translations. Vectors can be thought of
as displacements between two points (or multiples thereof) and are not affected by translations. Moreover, vector components can also
carry arbitrary dimensions ("units") as described in \cref{sec:units}, while point coordinates are necessarily lengths. 
Allowed operations follow the rules of the affine structure of Euclidean space: Points and length vectors can be added to return
another point. Subtracting two points yields a length vector. Scalar and cross-products are defined for vectors, respecting their
dimensions.

\subsubsection{Particle ID and physical properties}\label{sec:particles}
To distinguish different particle species, an integer \emph{particle ID} is introduced. It is used not only to label
a certain particle, but it also serves as an index to lookup tables containing information such as masses, lifetimes, electric charges, etc. These data are obtained at build time from the \texttt{particle}~\cite{Rodrigues_Particle} Python package, which provides a curated and up-to-date collection of particle properties derived from the PDG tables. At the same time, the CORSIKA-8-internal
particle IDs are generated for each particle by simply incrementing a counter, resulting in numbers currently in the range
\numrange{1}{231}. The numeric values never need to be used directly, though. Instead, particle IDs are exposed via human-readable \lstinline{enum class}es,
like \lstinline{Code::MuPlus} or \lstinline{Code::SigmaMinusBar} for higher expressiveness.

Conversion functions between \cor8 IDs and the \emph{Monte Carlo Particle Numbering Scheme} of the Particle Data Group
(PDG codes)~\cite{ParticleDataGroup:2022pth} are provided for interoperability with other software. These cannot conveniently
be used directly as internal particle IDs themselves because they sparsely cover a large range of numbers, making them unsuitable as lookup indices.

\subsubsection{Random-number generation}\label{sec:rng}
Being a Monte Carlo code, \cor8 relies heavily on (pseudo-)random numbers, which are used in large quantities. We have addopted the counter-based random number generator (CBRNG) architecture, and choose the Philox~\cite{Salmon:2011} as the default~\cite{Alves:2021pkr} algorithm. The defining
property of CBRNG is that their internal state is implemented as incrementable and decrementable counter, which allows advancing the state
by an arbitrary number of steps, at constant computational cost, to jump directly to a specific position in the sequence without passing through all intermediate numbers. This makes CBRNGs attractive for use in parallel algorithms that require random numbers, since a CBRNG-based iterable stream of random numbers can be broken into several independent sub-streams by dividing the counter range into many distinct nonoverlapping intervals, each still containing more numbers than the overall simulation needs.

When interfacing FORTRAN-based hadronic interaction models, \cor8 follows the same idea as \cor7 (which itself relies on the RNG RANMAR~\cite{James:1988vf,Marsaglia:1990}) of employing a distinct stream of random numbers for each module, each with its own seed. In particular, individual streams are forwarded to the hadronic interaction models. Technically, random numbers for the FORTRAN-based hadronic interaction models are generated in batches: A buffer is filled with 16 standard random numbers, to be consumed one at a time. Once the buffer is depleted, it is refilled to full capacity at once. This procedure drastically reduces the overhead of function calls from the FORTRAN code into our framework, which would otherwise occur abundantly and negatively impact performance.

\subsection{Environment}\label{sec:environment}
\cor8 is designed to allow the propagation of particle cascades in user-defined environments, providing a great deal of flexibility in three aspects:
\begin{itemize}
\item The medium of propagation can be freely selected, considering the limitations set by the physics modules.
\item Worlds consisting of different media can be modeled. Example use-cases include air showers that continue their
propagation below ground, e.g.\ in soil, water, or ice, or $\nu_{\tau}$-induced showers emerging from mountains. Transitions
between two media are also considered.
\item The environment/medium properties can be customized. Depending on the use-case, simulations may, e.g., require querying
(electro-)magnetic field strengths or the medium's refractive index as a function of position.
\end{itemize}

\subsubsection{Worldbuilding with the volume tree}
To simulate particle cascades in multiple media, a data structure and the corresponding algorithms must be set up.
This structure must map spatial regions to media in a flexible way that does not significantly impact performance. In \cor8 this is achieved with the \emph{volume tree}. The nodes of this tree structure are geometric primitives (volumes), such as spheres, and their placement in the tree represents geometric containment, i.e., a parent node contains its children. This arrangement makes it is possible to determine to which node a certain point belongs. More importantly,
the search space for calculating the possible entry and exit points of particle trajectories that cross volume boundaries is small because only a subset of all nodes needs to be considered.

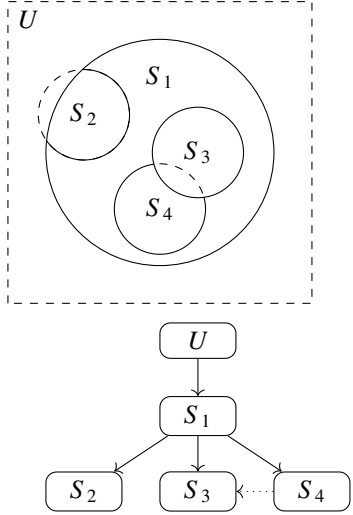
\begin{figure}
\centering
\begin{subfigure}[c]{5cm}
\begin{tikzpicture}[reverseclip/.style={insert path={(-99cm,-99cm) rectangle (99cm,99cm)}}]
\draw[dashed] (-2cm,-2cm) rectangle +(4cm,4cm);
\draw (-2cm,2cm) node[anchor=north west] {$U$};
\draw[dashed] (-1,.5) circle [radius=0.6cm];
\draw[clip] (0,0) circle [radius=1.5cm];
\draw (0,1cm) node {$S_1$};
\node at (-1.0,.5) [minimum size=1.2cm,shape=circle,draw] {$S_2$};
\node at (.5,0) [minimum size=1.2cm,shape=circle,draw] {$S_3$};
\begin{scope}
    \clip (.5,0) circle [radius=.6cm];
    \draw[dashed] (0,-.75) circle [radius=.6cm];
\end{scope}
\clip[reverseclip] (.5,0) circle [radius=.6cm];
\node at (0,-.75) [minimum size=1.2cm,shape=circle,draw] {$S_4$};
\end{tikzpicture}
\end{subfigure}
\par\medskip
\begin{subfigure}[c]{4cm}
\begin{tikzpicture}[every node/.style={rectangle,draw,minimum width=1cm,rounded corners},
    edge from parent/.style = {draw, ->},
    level distance=1cm,
    node distance=1cm
]
\node (U) {$U$}
child { node (S1) {$S_1$}
  child { node (S2) {$S_2$} }
  child { node (S3) {$S_3$} }
  child { node (S4) {$S_4$} }
};

\draw[->,dotted] (S4) -- (S3);
\end{tikzpicture}
\end{subfigure}
\caption{Example volume tree}
\label{fig:volumetree}
\end{figure}

Figure \ref{fig:volumetree} depicts an example world consisting of a number of spheres, $S_i$, with its corresponding volume tree.
The root node $U$, called \emph{Universe}, is equivalent to a sphere with infinite radius. It exists to make sure each point in space
can be mapped to a well-defined volume node. Complications arise when volumes intersect so that the idea of
containment must be reconsidered. We distinguish between two classes of such intersections:
\begin{itemize}
\item An intersection of a child volume with its parent ($S_1 - S_2$ in \cref{fig:volumetree}). In this case, the outer volume (parent) clips
the inner volume (child).
\item An intersection of two child nodes having the same parent requires additional information. For these
cases, nodes contain a (by default empty) list of pointers to \emph{excluded} nodes. A point does not belong to a node if it is contained by
any of the excluded nodes. In \cref{fig:volumetree} this is the case for the overlap between $S_3$ and $S_4$. $S_4$ has $S_3$ in its exclusion list (indicated
by the dotted arrow) so that points in the overlapping region belong to $S_3$.
\end{itemize}

For applications requiring observation surfaces of arbitrary shape, such as mountain-based observatories like the Tau Air-Shower Mountain-Based Observatory~\cite{TAMBO:2025jio} or muography studies with complex geometries, \cor8 will soon provide a triangular mesh geometry framework. This framework supports geometries loaded from standard \emph{Wavefront OBJ}~\cite{wavefront-obj} or binary \emph{PLY}~\cite{ply-format} files, which must be oriented and watertight to be effectively integrated into the volume tree. To maintain performance for large meshes, ray–triangle intersections are accelerated using a bounding volume hierarchy that recursively partitions the triangles into a binary tree of axis-aligned bounding boxes, reducing the average intersection query cost from linear to logarithmic in the number of triangles.

\subsubsection{Dressing volumes with models}
The individual volumes alone do not constitute any description of the media and their physical properties. Therefore, they have to be furnished
with models of these properties. For this purpose, we employ dynamic polymorphism: An abstract class defines the interface and serves as base class. Implementations
of this interface occur in classes that inherit from the interface class. A special challenge arises in \cor8 because the medium interface is
not fixed and depends on the physics modules one wants to use for a particular simulation: For example, Cherenkov light and radio emission modules need to query
the index of refraction while electromagnetic interaction models require material constants like the radiation length, and
for studies of EAS development in thunderclouds, electric fields need to be modeled. To accommodate these requirements, we make use of \emph{mixin inheritance},
which is a template class that inherits from its template argument (see e.g.\ Ref.~\cite{Nesteruk:2018}). This way, several interface classes,
each one responsible for a single aspect, can be chained together to form the complete interface. \Cref{lst:environmentMixin} shows a simplified
example of how three different properties are combined to form a single abstract base class.

\begin{figure*}
\begin{lstlisting}[label=lst:environmentMixin,caption={Mixin-based environment interface composition (simplified)}]
struct IMediumModel {
  virtual DensityType getMassDensity(Point const &) const = 0;
};

template <typename T> struct IRefractiveIndexModel : public T {
  virtual double getRefractiveIndex(Point const &) const = 0;
};

template <typename T> struct IMagneticFieldModel : public T {
  virtual MagneticFieldVector getMagneticField(Point const &) const = 0;
};

using MediumInterface = IMagneticFieldModel<IRefractiveIndexModel<IMediumModel>>;
\end{lstlisting}
\end{figure*}

The actual implementations of the interfaces, in the end required to inherit from their respective interface class, follow the same pattern so that they
can be freely combined while orthogonal aspects stay independent. \Cref{lst:environmentMixinImpl} shows example implementations of the interfaces of \cref{lst:environmentMixin} and their
usage via dynamic polymorphism. Two objects, \lstinline{modelA} and \lstinline{modelB}, are created each containing a different implementation of the \lstinline{getMassDensity()}
interface. The function \lstinline{printMassDensity()} handles both objects via their interface, the \lstinline{MediumInterface} class from \cref{lst:environmentMixin},
and is agnostic about the implementation, which is selected only at runtime.
Similarly, volume nodes are linked to models only via their interfaces.

The most fundamental medium properties needed in \cor8 are density and the fractions of the isotopes. They always need
to be specified even for the most basic cascade simulations.

\begin{figure*}
\begin{lstlisting}[label=lst:environmentMixinImpl,caption={Mixin-based composition of implementations (simplified)}]
template <typename T> struct FlatExponentialDensity : public T {
  // [...]

  virtual DensityType getMassDensity(Point const &P) const override {
    return rho0 * exp(axis.dot(P - Pref) / scaleHeight);
  }
};

template <typename T> struct HomogeneousDensity : public T {
  // [...]

  virtual DensityType getMassDensity(Point const &p) const override {
    return rho0;
  }
};

template <typename T> struct ExponentialRefractiveIndex : public T {
  // [...]

  virtual double getRefractiveIndex(Point const &p) const override {
    // [...] some implementation
  }
};

template <typename T> struct UniformMagneticField : public T {
  // [...]

  virtual MagneticFieldVector getMagneticField(Point const &p) const override {
    // [...] some implementation
  }
};

void printMassDensity(MediumInterface const& medium) {
  Point const p = make_some_point(); // obtain a Point
  
  // query density at point p
  // concrete implementation selected via dynamic dispatch
  DensityType const rho = medium.getMassDensity(p);
  std::cout << "density at p = " << rho << std::endl;
}

int main() {
  ExponentialRefractiveIndex<
      FlatExponentialDensity<UniformMagneticField<MediumInterface>>>
      modelA;

  ExponentialRefractiveIndex<
      HomogeneousDensity<UniformMagneticField<MediumInterface>>>
      modelB;

  printMassDensity(modelA); // calls FlatExponentialDensity::getMassDensity()
  printMassDensity(modelB); // calls HomogeneousDensity::getMassDensity()

  return 0;
}
\end{lstlisting}
\end{figure*}

\subsection{Processes}
Processes are the entities that act on particles in various ways. They are grouped in six categories:

\begin{description}
\item[InteractionProcess] This class of processes models interactions and related functionality. Typical examples are wrappers around hadronic interaction models.
Two methods must be provided: \lstinline{CrossSectionType getCrossSection(TParticle const&, Code const, FourMomentum const&)} needs to return the (possibly infinite)
interaction length of the modelled physical process for the particle being propagated in its current medium. The actual event generation has to be performed in the \lstinline{doInteraction(TSecondaryView&, Code const, FourMomentum const&)}
method, which typically adds secondary particles via the \lstinline{SecondaryView} object.

\item[DecayProcess] This class of processes models decays. Methods to be provided are: \lstinline{TimeType getInverseLifeTime(Particle const&)}, which returns
the lab-frame decay time of the particle being propagated. Analogous to \lstinline{InteractionProcess}, the decay event is generated in the \lstinline{doDecay(SecondaryView&)} method,
which typically fills the decay products into the \lstinline{SecondaryView}.

\item[BoundaryCrossingProcess] This class of processes is relevant if any action shall be performed when a particle exits its current volume to enter
another. In that case, the method \lstinline{doBoundaryCrossing(Particle&, VolumeNode&, VolumeNode&)} is called. The last two parameters
refer to the current and new volume nodes, respectively.

\item[SecondariesProcess] After an interaction or decay event has been performed, the newly generated secondaries can be processed further, e.g.\ filtered
and/or modified. This happens in the \lstinline{doSecondaries(StackView&)} method, in which the secondaries are typically iterated over.

\item[ContinuousProcess] In this class of processes, aspects concerning the continuous movement of a particle along its trajectory are handled. Before the
actual propagation takes place, its \lstinline{LengthType getMaxStepLength(Particle const&, Trajectory const&)} method is called, which returns a maximum (possibly infinite) step-length.
It serves as a hint to the propagation to limit the step-length to that value if necessary. This functionality is provided to ensure numerical accuracy. For instance, a process implementing
energy losses may limit the step-length to make sure that the energy of the particle does not change too much so that the decay time stays approximately
valid. \lstinline{ProcessReturn doContinuous(Step<TParticle>&, bool const)} is executed after the length of the trajectory has been determined.
The particle properties may be modified, e.g.\ the energy reduced. The boolean input parameter indicates whether the previous step-length limitation
had been caused by the \lstinline{getMaxStepLength()} method of the same process.

The process can indicate whether the particle shall be regarded as absorbed via the return value, which is an \lstinline{enum} called \lstinline{ProcessReturn}.
\item[StackProcess] Primarily for statistical purposes, it is beneficial to execute code periodically after each $N$ cascade steps. This functionality
is provided with this class of processes, which require a \lstinline{doStack(Stack&)} method. An example use-case is the estimation of the remaining runtime of
the simulation by checking how much of the initial energy is still stored in particles on the stack. This quantity decreases linearly with time and reaches zero at the end of the run.
\end{description}

An arbitrary number of processes can be combined to constitute the \emph{process sequence}.

\subsection{Particle stack}\label{sec:stack}
The particle stack is the central object that manages the particles in memory. The current implementation uses a \emph{structure of arrays}-like
memory layout for the particles, meaning that each particle property is stored in a separate, contiguous array.
This also means that no independent, compact ``particle object'' that one could create anywhere exists. Instead, a ``particle'' is a mere reference
to the actual data in the individual arrays. This reference object provides methods like \lstinline{particle.getEnergy()}, \lstinline{particle.setPID(Code)},
etc., so that the developer can be oblivious to the underlying memory layout. 

The default particle properties stored on the stack are:
\begin{itemize}
\item the particle ID (\lstinline{Code}, integer, also containing nuclear isotope data ($A,Z$)),
\item its kinetic energy (\lstinline{HEPEnergyType}),
\item its position (\lstinline{Point}),
\item its direction vector (dimensionless, normalized \lstinline{Vector}),
\item the time (\lstinline{TimeType}),
\item a pointer to current volume,
\item a boolean flag indicating whether a particle is deleted.
\end{itemize}

To avoid searching the volume tree each time the current volume is needed, a pointer is stored. It is updated each time a boundary is crossed, as described below.

The \emph{is-deleted} flag allows a particle to be marked as ``removable'' at any position on the stack. A marked particle is only removed when it is read again on top of the stack. This procedure is useful e.g.\ for thinning and similar filtering purposes in-place, without the need for a (costly) reordering of the particles. It is also a cornerstone of the cascade history described in \cref{sec:cor8_lineage}.

\subsection{Program flow}\label{sec:flow}
We are now in a position to discuss how the individual building blocks gear into each other to process a full
particle cascade. The basic principle is not different from the standard prescription of Monte Carlo cascade codes: particles from
the stack are propagated and if they produce any secondaries these are pushed onto the stack. In \cor8, this main loop
is implemented in the \lstinline{Cascade} class, which has access to the stack, the environment and the process sequence.
Furthermore, one may select a certain tracking algorithm, i.e., the procedure by which the trajectory is determined. Propagation in magnetic fields is much more complex than propagation without magnetic fields.
A propagation step for a single particle consists of four parts: first, the trajectory is calculated, followed by determining the step-length to be taken. Then, the particle is propagated by that length. Finally, the action corresponding to the chosen step-length is executed. Let us consider each of these substeps in more detail.

\subsubsection{Trajectory determination}\label{sec:trajDeterm}
Without a Lorentz force, the trajectory is a straight line (ray). The only calculation necessary is determining its maximum
length, which is determined by its intersection with boundaries of the volume. In the volume tree, only a subset of the
existing volumes needs to be considered: the current volume itself, its child volumes, and its excluded volumes.
At the time of writing, the only implemented geometric primitive is a sphere. For the intersection of a straight line parameterized by a real number, $\ell$, such that $\ell = 0$ corresponds to the current position and $\ell > 0$ ($\ell < 0$) corresponds
to points in front of (behind) the current position, 
with a sphere, a quadratic equation must be solved. This equation yields up to two real solutions for $\ell$. Care must be taken to
choose the correct solution: Negative solutions are always excluded because they refer to the past. If both solutions are positive,
the smaller (greater) value must be chosen when entering (exiting) a sphere.

A complication arises when a particle has moved
to the boundary in the previous step. Due to limited numerical precision, the particle appears to be slightly before or after
the boundary. If the intersection is recalculated in the current step and the particle appears to be still before the boundary,
the same crossing can be proposed again. This results in a stuck particle and an infinite loop. This can be avoided by keeping
a reference to the current node in memory and updating it during a boundary crossing. This ensures that it is always clear whether the considered volume is being entered or exited.

\begin{figure}
    \centering
    \begin{tikzpicture}[
    >=Latex,
    point/.style={fill=black,rectangle,minimum size=5pt,inner sep=0pt},
    vec/.style={->,line width=1.4pt},
    path/.style={line width=0.8pt},
    dashedcurve/.style={dashed,line width=1pt}
]

\coordinate (xn) at (0,0);
\coordinate (xmid) at (5,2);
\coordinate (xnpone) at (8,-1);

\draw[path] (xn) -- (xmid) -- (xnpone);

\node[point] at (xn) {};
\node[point] at (xmid) {};
\node[point] at (xnpone) {};

\draw[vec] (xn) -- ($(xn)!0.4!(xmid)$)
    node[above left] {$\vec{u}_0$};

\draw[vec] (xmid) -- ($(xmid)!0.4!(xnpone)$)
    node[right] {$\vec{u}_1$};

\node[below left] at (xn) {$\boldsymbol{P}_0$};
\node[above] at (xmid) {$\boldsymbol{P}'$};
\node[below right] at (xnpone) {$\boldsymbol{P}_1$};

\draw[dashedcurve]
    (xn)
    .. controls ($(xn)!0.6!(xmid)$) 
       and ($(xnpone)!0.6!(xmid)$)
    .. ($(xnpone)+(-0.2,-0.1)$);

\end{tikzpicture}
    \caption{Schematic of the leap frog algorithm.}
    \label{fig:leap_frog}
\end{figure}
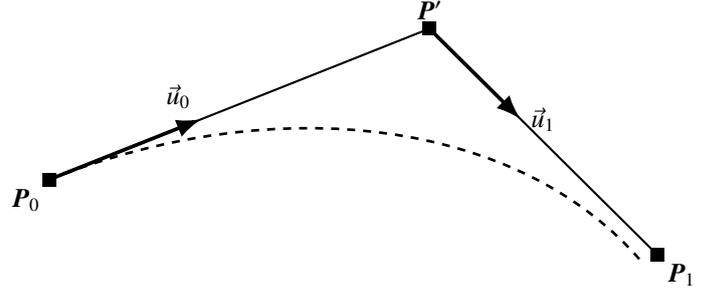

The situation with magnetic fields is more involved. In this case, we employ the leapfrog algorithm as implemented in AIRES and described
in Ref.~\cite{Cillis:1997cr}. Updating the position from $\mathbf P_0$
to $\mathbf P_1$ consists of two half steps, between which the direction $\vec u$ is updated as shown in \cref{fig:leap_frog}:
\begin{align}
\mathbf P' &= \mathbf P_0 + \frac{\ell}{2} \vec u_0, \\
\vec u_1 &= \vec u_0 + \vec u_0 \times \ell \frac{q}{p} \vec B, \\
\mathbf P_1 &= \mathbf P' + \frac{\ell}{2} \vec u_1.
\end{align}
If we consider the magnetic field $\vec B$ constant during the propagation (we set $\vec B = \vec B(\mathbf P_0)$) and
do not normalize $u_1$, then $\mathbf P_1(\ell)$ is a parabola.
The intersection with a sphere is described by a quartic equation that can be solved analytically, albeit with greater computational effort.
The correct solution to select is the one with the smallest positive value. Additionally, several safeguards are introduced to handle situations in which the procedure proves to be numerically unstable.

For charged leptons, multiple Coulomb scattering is taken into account following the treatment by Moli\`{e}re \cite{Moliere:1948zz} as implemented in PROPOSAL.
The effect of multiple scattering is implemented as an update of the direction vector; the additional translation of the position vector is currently neglected.
In PROPOSAL, also stochastic deflections in electromagnetic interactions are implemented \cite{Gutjahr:2022quk}, but this subdominant effect is currently ignored for \cor8.

\subsubsection{Step-length determination}\label{sec:stepLength}
After the trajectory is proposed, the distance that the particle will  propagate is determined. To do so, a number of candidate step-lengths
are considered, out of which the minimum is chosen:
\begin{itemize}
\item The maximum step-length to reach a volume boundary as described in the previous section.
\item In case of magnetic fields, the angular deflection per step is limited to a user-configurable value. The default value is \SI{0.2}{rad}.
\item A candidate time of decay is sampled from the current (lab-frame) lifetime and converted to a length using
the current velocity. The current lifetime is queried by summing the contributions (branching ratios) of all \lstinline{DecayProcess}es in the process sequence:
\begin{equation}
  \frac{1}{\tau_{\mathrm{tot}}} = \sum_i \frac{1}{\tau^{(i)}}.
\end{equation}
\item A candidate interaction point is sampled: First, a grammage is sampled from the exponential distribution with
the current interaction length $\lambdaInt$ as parameter, which is determined from the contributions of all (competing) \lstinline{InteractionProcess}es in the process sequence:
\begin{equation}
  \frac 1 {\lambda_{\mathrm{int,tot}}} = \sum_i \frac{1}{\lambdaInt^{(i)}}.
\end{equation}
To accomplish this, the individual \lstinline{InteractionProcess}es themselves typically calculate the sum of the cross-sections for each isotope in the medium of the current volume, weighted by its corresponding fractional abundance. Then, the grammage is converted to length using a function provided by the density model. Except
for the trivial case of homogeneous density, the density models usually assume a rectilinear trajectory.
\item The minimum of the values obtained from the \lstinline{getMaxStepLength()} methods of the \lstinline{ContinuousProcess}es.
\end{itemize}

\begin{figure}
    \centering
    \includegraphics[width=\linewidth]{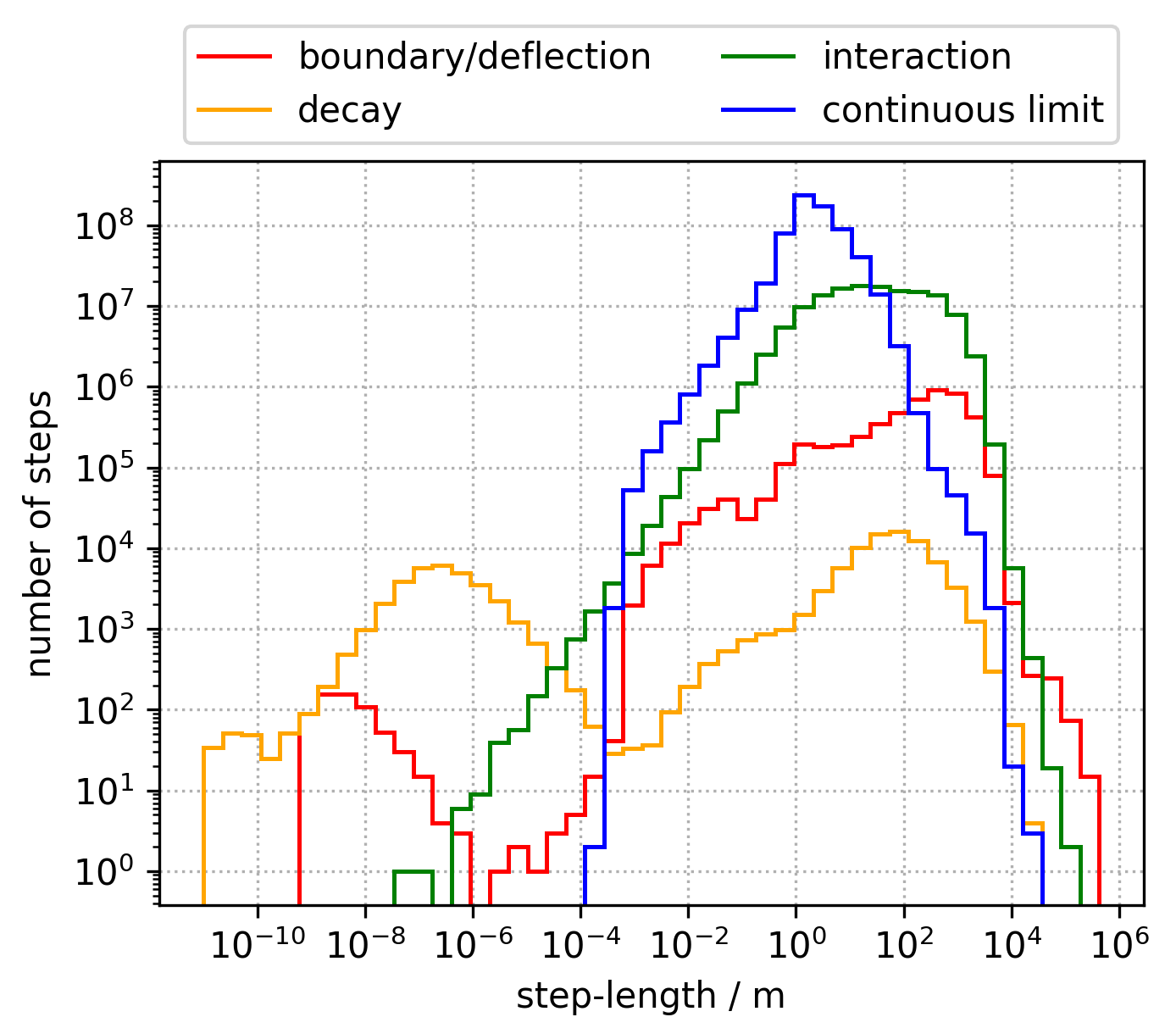}
    \caption{Distributions of step lengths, categorized by the process providing the minimum candidate step length.}
    \label{fig:step_length}
\end{figure}

Each step can be classified according to the candidate step length that determined the minimum. Example distributions of step lengths, accumulated from 100 proton-induced vertical air showers with a primary energy of \SI{10}{TeV}, are shown in \cref{fig:step_length}. These distributions offer useful diagnostic insight into the propagation algorithm.

\subsubsection{Propagation}\label{sec:prop}
Once the step-length has been selected, the initially proposed trajectory is shortened to that length. Then, continuous processes
are applied to the particle along the trajectory, which may modify its properties. Purely ``observing'' continuous processes, which do not modify the particle properties, are also
possible. Examples include calculating the radio emission and recording the longitudinal profile.

The step-length is additionally constrained by energy-loss considerations such that at most 10\,\% of the particle's energy is lost over a single step or such that the particle energy drops just below the energy cut at the end of the step. In the latter case, it is absorbed and marked for deletion from the stack.

\subsubsection{Final action}
In case a particle survives the continuous processes, the action corresponding to the selected minimum step-length is performed:
\begin{itemize}
\item If the step-length was set by a step-length limitation of a continuous process, nothing happens. All steps described in \cref{sec:trajDeterm,sec:stepLength,sec:prop}
will be repeated.
\item In case of an interaction, the interaction length is recalculated based on the particle properties after the propagation, denoted $\lambda_{\mathrm{int,tot}}'$. This
may differ from the initial value $\lambda_{\mathrm{int,tot}}$, which was also used for the sampling. One of the competing
\lstinline{InteractionProcess}es needs to be selected to perform the interaction. This is done by randomly selecting the $i$-th process with probability
$p_i = \max(\lambda_{\mathrm{int,tot}}, \lambda_{\mathrm{int,tot}}') / \lambda_{\mathrm{int}}^{(i)}$. If $\lambda_{\mathrm{int,tot}} < \lambda_{\mathrm{int,tot}}'$, the
probabilities will not sum up to one. The remaining fraction corresponds to a rejection of the interaction. 

The sampling in this case is still exact despite the change of $\lambda_{\mathrm{int,tot}}$. If, on the other hand, $\lambda_{\mathrm{int,tot}} > \lambda_{\mathrm{int,tot}}'$, the sampling
is in principle not exact, but for sufficiently small changes in energy the incurred error is negligible.
\item For decays, the procedure is analogous.
\item In case of a volume boundary transition, the \lstinline{doBoundaryCrossing()} methods of all \lstinline{BoundaryCrossingProcess}es are called, in which
the particle may be marked for deleted. Additionally, the pointer to the current volume is updated to the new volume.
\end{itemize}

Interactions and decays create new secondaries, which are pushed onto the stack via the \lstinline{SecondaryView} that offers an \lstinline|addSecondary()|
method, which is called with the properties of the secondary. After the \lstinline{SecondaryView} has been filled, it is passed subsequently to all \lstinline{SecondariesProcess}es.
These may iterate over the new secondaries and alter or even mark as delete some of them again. Finally, the projectile particle, now no longer on top of the stack, is marked as deleted.

\subsection{Output infrastructure}
\cor8 produces a wide range of output data, from compact datasets such as longitudinal profiles to very large particle samples recorded at the observation level. Correspondingly, typical use cases range from single ultra-high-energy cosmic-ray showers to large ensembles of gamma-ray showers with comparatively smaller per-shower output. For reference, the simulations used in \cref{sec:hadronic_showers} typically produce output files of about 100\,MB per shower.

At the time of writing, the simulation output is written using a combination of the Apache \emph{Parquet}~\cite{parquet} format and accompanying \emph{YAML} files. Parquet is used to store the binary event data in a columnar layout that enables efficient I/O and is widely supported across programming languages. The YAML files provide a human-readable representation of the simulation configuration and a description of the output content. Each output module writes its data independently, and all files are organized in a hierarchical directory structure. A special case in the current implementation are the interaction histograms. They are implemented using the \emph{Boost.Histogram}~\cite{Schreiner:2020} library, which supports multi-dimensional histograms with flexible axis definitions. Since Boost.Histogram does not define a native persistent storage format, the histograms are currently saved using a lightweight \emph{NumPy}-based file format~\cite{Harris:2020xlr}.

We provide a Python library for data access and analysis, that allows for the convenient reading and processing of the generated output~\cite{corsika8_python}, thereby making integration into user codes and analysis workflows transparent and independent of the underlying \cor8 output file format. We strongly recommend that users build on the provided read-in library because the underlying file format might change in the future if needed. However, should lower-level integration be strictly necessary, Parquet readers are readily available for many programming languages.

\begin{figure*}
    \centering
    \includegraphics[width=\linewidth]{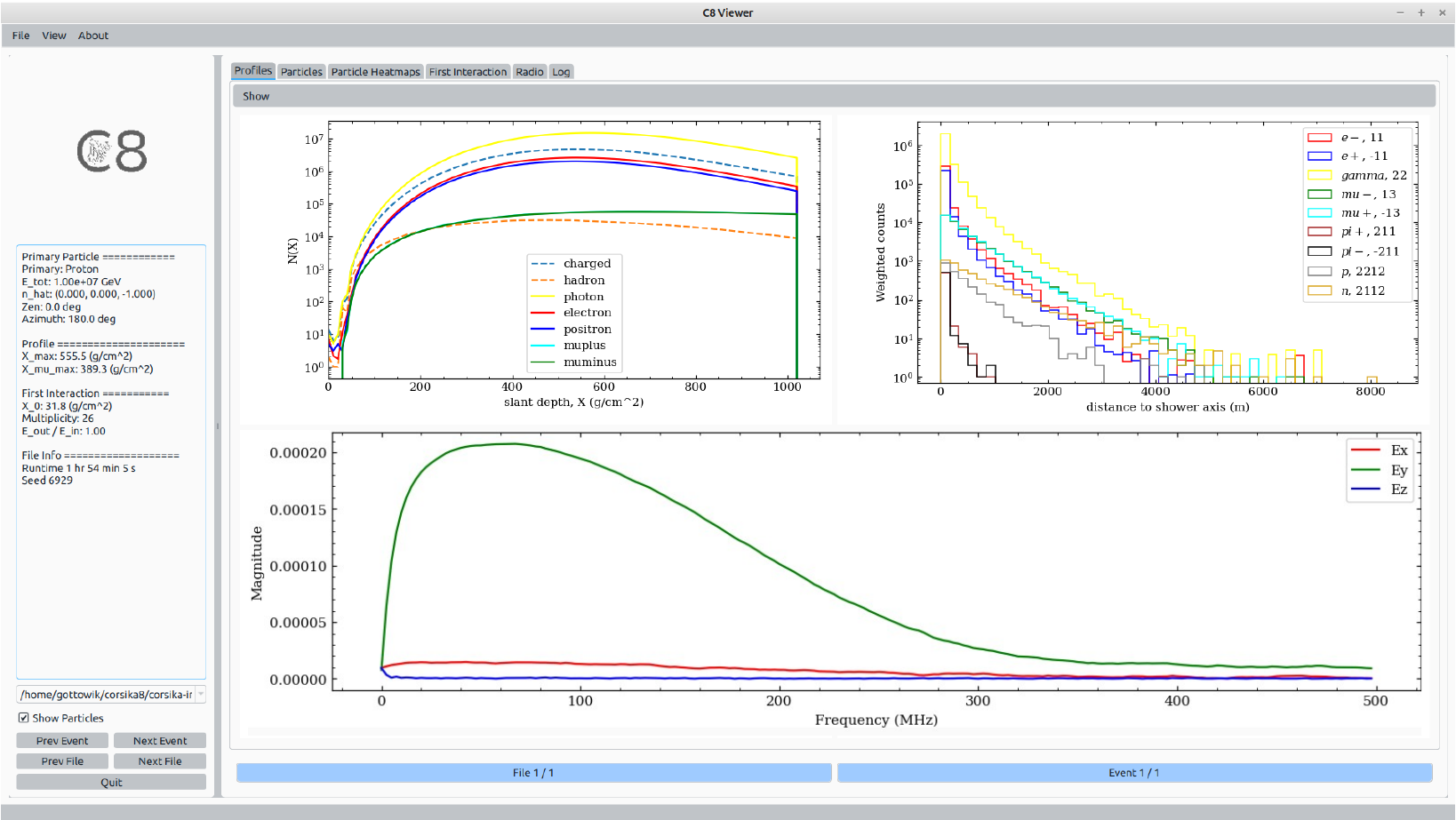}
    \caption{Collection of visualizations from the \emph{\cor8 Viewer}. The top-left panel shows the longitudinal particle profile of the shower, while the top-right panel displays the particle distribution at ground level. The bottom panels present the radio frequency spectrum evaluated at a selected observer position.}
    \label{fig:c8_viewer}
\end{figure*}

Furthermore, the \emph{\cor8 Viewer}~\cite{corsika8_gui}, shown in \cref{fig:c8_viewer}, is provided as a graphical user interface (GUI) for interactive inspection and visualization of simulated \cor8 showers. The viewer allows users to explore the particle content, longitudinal profiles, and radio emission at ground level directly from the simulation output. This offers an intuitive way to validate simulations and inspect individual events.

\subsection{Atmosphere}
A key component of all air shower simulations are models of the atmosphere, particularly its density, $\varrho(h)$, as a function of altitude, $h$, which often consists of multiple layers.
One well-known model is Linsley’s parameterization of the U.S.\ Standard Atmosphere~\cite{USStandardAtmosphere1976}, which consists of four isothermal-barotropic layers (i.e., exponential decrease of density with altitude with a fixed scale height) and one topmost layer with a constant density. Such layered parameterizations provide a compact and analytically tractable description of the average atmospheric density profile and have therefore been widely used in air-shower simulations for a variety of applictions.

\cor8 follows this general approach, but implements the atmosphere within a modular and extensible framework. Rather than using Linsley's original parameterization, described e.g.\ in Ref.~\cite{Cruz-Moreno:2013}, we have adopted Keilhauer's parameterization~\cite{Keilhauer:2004jr} as the default option. This parameterization consists of the same number of layers but the scale heights and reference densities are different. Linsley's original parameterization exhibits discontinuities in density at the layer boundaries, which lead to unphysical artifacts seen in muon production profiles \cite{Reininghaus:2022qyl}. For convenience, the coefficients of all atmosphere parameterizations available in \cor7 are also provided. Furthermore, users are free to specify an arbitrary number of exponential and homogeneous layers.

For the simulation, it is technically more important to consider the integrated density along a particle's trajectory, called grammage, $X = \int \varrho \mathrm d l$, and its inverse rather than to consider the density $\varrho(h)$ itself.
Since the relationship between altitude and path length in a spherical atmosphere is nonlinear, the integration cannot be performed analytically using closed-form expressions.
The mathematical treatment of this integral in an isothermal-barotropic density model leads to the definition of the Chapman function~\cite{Chapman:1931}, on which a substantial number of approximations have been developed~\cite{Huestis:2001,Vasylyev:2021}.
The implementation in \cor8 follows the \emph{sliding planar atmosphere} approximation, which was originally developed for AIRES. It is also used in \cor7 when the \emph{curved} option~\cite{Heck:2004} is enabled, where it is applied locally during particle transport with a limitation on the step length to ensure the validity of the approximation. For the precalculation of the slant depth along the shower axis, however, \cor7 employs a separate numerical integration.
In \cor8, this approximation is used to compute the grammage along particle trajectories, while the particle propagation itself is performed in the full spherical geometry. In contrast to \cor7, no explicit step-length limitation is currently imposed, such that very long and highly inclined trajectories can lead to a reduced accuracy.
The main assumption is that the atmosphere is considered locally flat with an isothermal-barotropic density model having a scale height $\hScale$. In this case, the grammage along a rectilinear trajectory starting at point $\bm P_0$, pointing in the (normalized) direction $\vec u$ and having length $L$ can be expressed as
\begin{equation}
X = \varrho(\bm P_0) \frac{\hScale}{\vec a \cdot \vec u} \left(\exp\left(\vec a \cdot \vec u \frac{L}{\hScale}\right) - 1\right).
\label{eqn:XflatExp}
\end{equation}
Here, $\vec a$ denotes a unit vector pointing in direction of the gradient, i.e., vertically downwards, that gets recalculated for each trajectory as pointing from the starting point towards the center of the Earth. In contrast, in a globally flat model, $\vec a$ would be a fixed constant.

In contrast to \cor7, the atmosphere in \cor8 is treated as a generic propagation medium rather than a hard-coded special case. This allows for the implementation of different atmospheric descriptions in a uniform and consistent manner. This flexibility is part of \cor8{}’s extensible medium framework, allowing more general density profiles to be implemented when desired. The exponential model remains a commonly used baseline due to its simplicity and analytical tractability.

\section{Electromagnetic showers}

Having discussed the cornerstones of the \cor8 design, we will now showcase the simulation of electromagnetic particle showers in air.

\subsection{Implementation and results}
\begin{figure}
  \includegraphics[width=\linewidth]{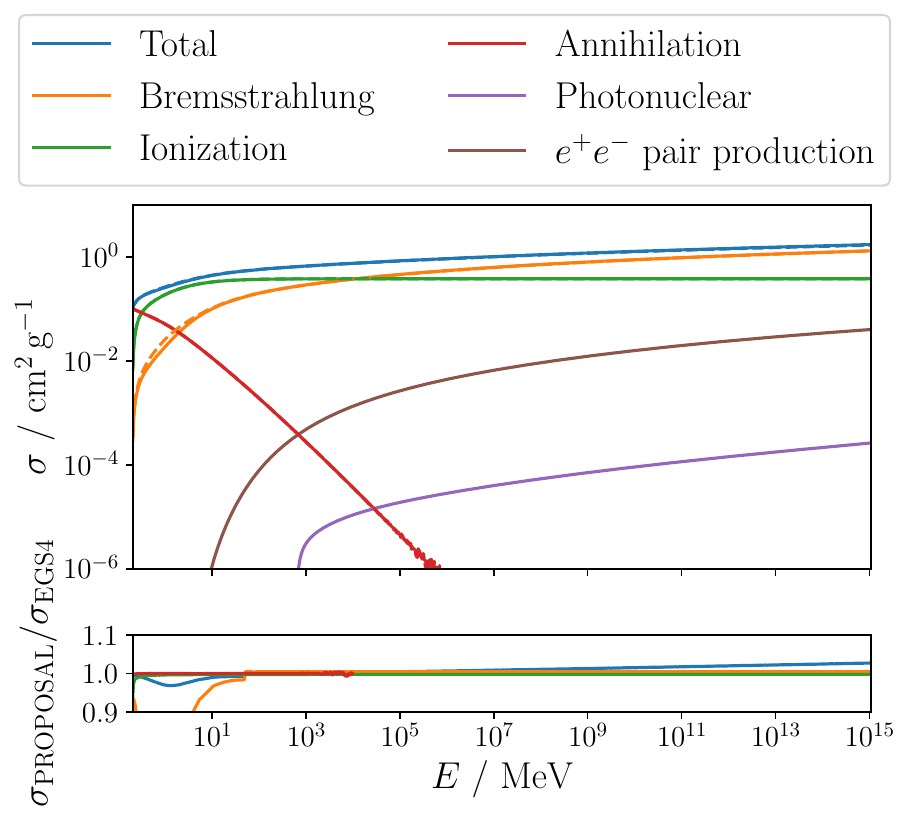}
  \includegraphics[width=\linewidth]{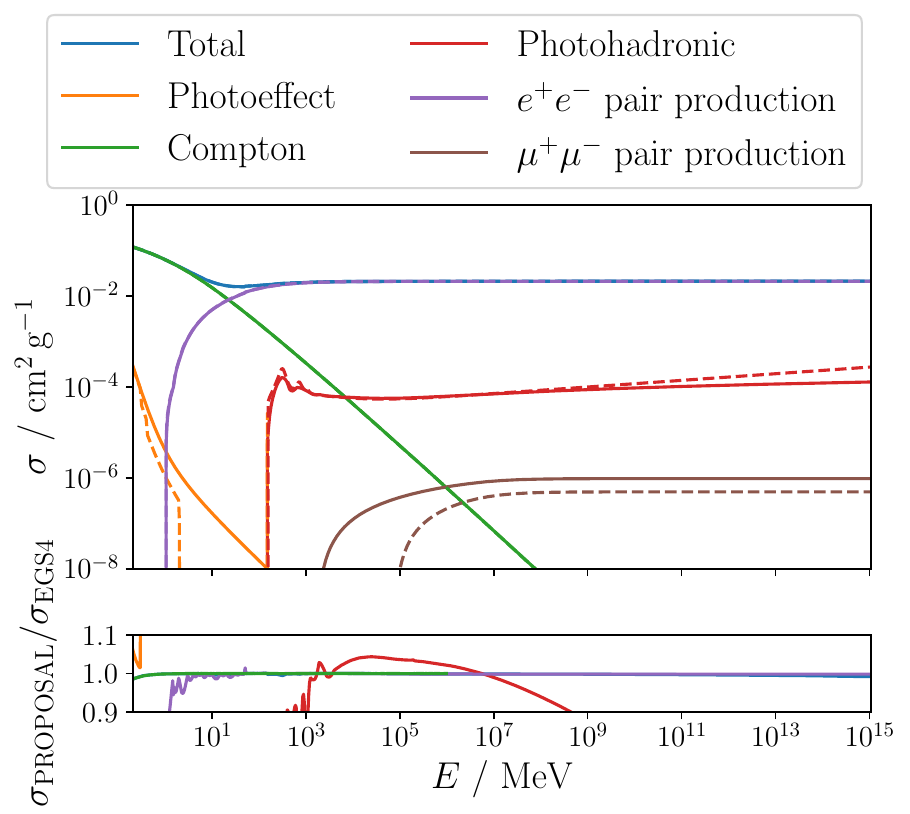}
  \caption{Total stochastic cross-sections for positrons (top) and photons (bottom) in air. The losses smaller than \SI{0.2}{MeV} are treated continuously to obtain a finite cross-section. The dashed lines refer to cross-sections from \cor7, the solid lines show the implementation in \cor8.}
  \label{fig:xsec}
\end{figure}

Earlier versions of CORSIKA simulated the electromagnetic component of air showers using a customized version of EGS4~\cite{Nelson:1985EGS4}, which was deeply integrated into the source code of CORSIKA.
Alternatively, the electromagnetic shower component could be described by the analytic NKG formul\ae~\cite{Kamata:1958xka,Greisen:1956}.
In \cor8, the electromagnetic component is currently being simulated by the lepton propagator PROPOSAL~\cite{Koehne:2013gpa,Dunsch:2018nsc,Alameddine:2020zyd}, which is also used to describe muon and tau lepton interactions.
To ensure comparability with the legacy version of CORSIKA, the cross sections from EGS4 were implemented in PROPOSAL~\cite{Alameddine:2023wrp}.
The standard parameterization of the bremsstrahlung cross section is taken from Ref.~\cite{Koch:1959zz}; above \SI{50}{MeV}, the extreme relativistic cross section with Coulomb correction is used, while at lower energies an empirical correction factor is added, which thus rescales the total cross section.
The cross section for electron-positron pair production by a photon uses the same screening functions as the bremsstrahlung cross section with a low-energy empirical correction factor taken from Ref.~\cite{Storm:1970ouq}.
The photoelectric effect parameterization was taken from Refs.~\cite{Sauter:1931a,Sauter:1931b} for the K$\alpha$ shell, along with an empirical correction factor from Ref.~\cite{Hubbell:1969xbg}.
The cross section for photohadronic interactions is identical to its description in \cor7~\cite{Heck:2012}, while interfaces to SOPHIA~\cite{Mucke:1999yb} and SIBYLL~2.3d~\cite{Riehn:2019jet} were implemented for the actual event generation. The implemented cross sections differ at most by a few percent, and only at the lowest or highest energies or in regions where the contribution of the process is very small, cf.\ \cref{fig:xsec} for exemplary comparisons of positron and photon cross sections between \cor7 and \cor8.
Lastly, PROPOSAL simulates electron-positron pair production by an ingoing electron or positron -- a process which was not taken into account by \cor7.

Unlike the highly integrated usage of EGS4 in legacy CORSIKA versions, PROPOSAL within \cor8 is integrated analogously to the hadronic interaction models. This allows for the addition of other electromagnetic interaction models alongside PROPOSAL in the future.
Furthermore, the modular structure of PROPOSAL allows for disabling, interchanging, or adapting of individual interaction parameterizations.

\begin{figure}
    \centering
    \includegraphics[width=0.48\textwidth]{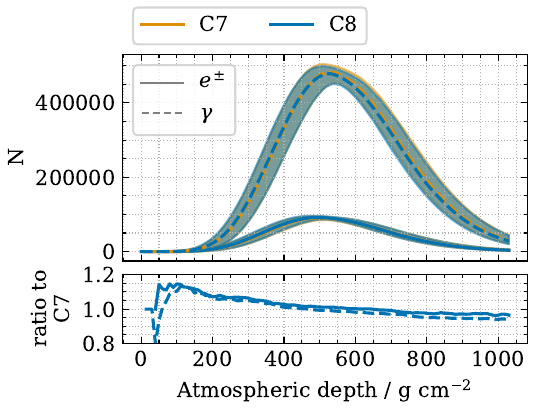}
    \caption{Median longitudinal profile of electromagnetic particles with energies above \SI{1}{MeV} in \SI{100}{TeV} photon-induced showers in \cor7 and \cor8. The shaded bands indicate the interquartile range.}
    \label{fig:long_100TeV}
\end{figure}

\begin{figure}
    \centering
    \includegraphics[width=0.48\textwidth]{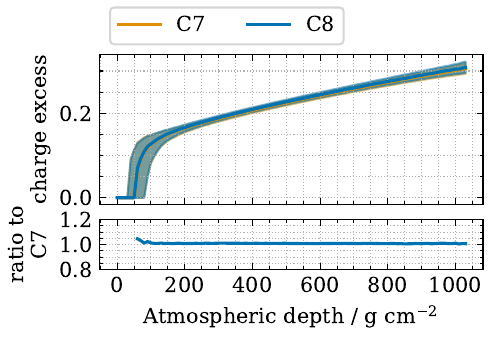}
    \caption{Median charge excess of \SI{100}{TeV} photon-induced showers in \cor7 and \cor8.}
    \label{fig:ch_exc_100TeV}
\end{figure}

\begin{figure}
    \centering
    \includegraphics[width=0.48\textwidth]{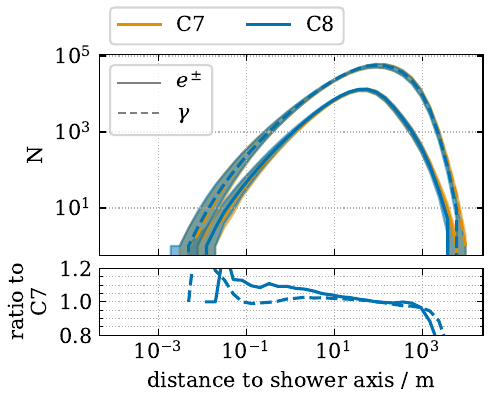}
    \caption{Median lateral profile of electromagnetic particles in \SI{100}{TeV} photon-induced showers at a height of \SI{5.8}{\kilo\meter}, typical depth of shower maximum, in \cor7 and \cor8.}
    \label{fig:lat_100TeV}
\end{figure}

\begin{figure}
    \centering
    \includegraphics[width=0.48\textwidth]{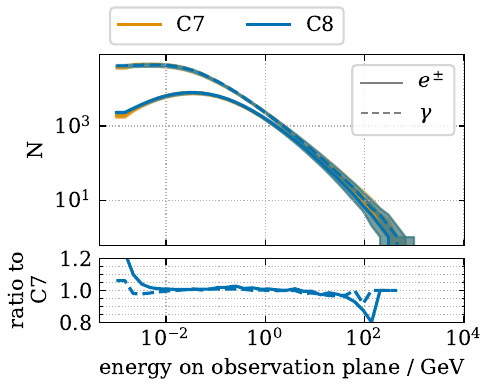}
    \caption{Median energy distribution of electromagnetic particles in \SI{100}{TeV} photon-induced showers at a height of \SI{5.8}{\kilo\meter}, typical depth of shower maximum, in \cor7 and \cor8.}
    \label{fig:E_100TeV}
\end{figure}

Comparisons of longitudinal and lateral profiles, as well as the charge excess, defined as $(N_{e^-}-N_{e^+})/(N_{e^-}+N_{e^+})$, of electromagnetic showers at a primary energy of \SI{100}{TeV} with \cor7 are shown in \cref{fig:long_100TeV,fig:ch_exc_100TeV,fig:lat_100TeV,fig:E_100TeV}. Further information on the exact configuration is provided in \cref{sec:mc_config}. In all profile plots, the solid lines show the median value and the shaded bands indicate the interquartile range (25th to 75th percentile), representing shower-to-shower fluctuations. For this comparison, a total of \num{2500} photon-induced showers were simulated with \cor8 and \cor7. The agreement between \cor7 and \cor8 is on the few-percent level, except for very low particle numbers at the earliest stage of the shower development and very small distances to the shower axis.


\begin{figure}
  \centering
  \includegraphics[width=0.48\textwidth]{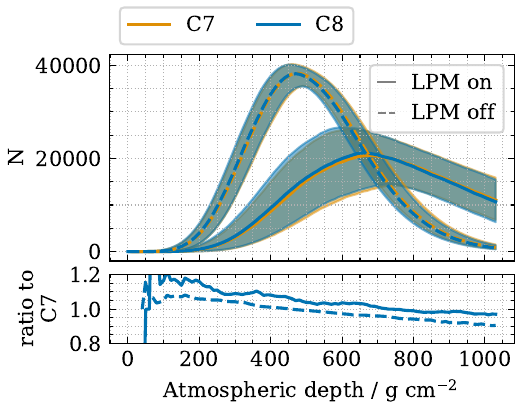}
  \caption{Median longitudinal profile of charged particles above \SI{100}{TeV} in \SI{100}{EeV} photon-induced showers with and without the LPM effect in \cor7 and \cor8.}
  \label{fig:long_100EeV}
\end{figure}

The Landau-Pomeranchuk-Migdal (LPM) effect~\cite{Landau:1953um,Migdal:1956tc} changes the behavior of electromagnetic showers at extremely high energies and at large atmosphere densities. Under these circumstances, the approximation of an interaction with a single isolated atom breaks down and the coherence of the initial and final wave functions is lost due to scattering on other atoms.
This suppresses the emission of bremsstrahlung photons, particularly those with a low energy compared to the initial lepton energy, as well the production of electron-positron pairs, particularly those with a symmetric energy distribution between the produced leptons. This effect slows down the development of the cascade and shifts the shower maximum to deeper depths.  For earlier work on electromagnetic cascades in water and lead and the formul\ae{} used to calculate the LPM effect, we refer the reader to Ref.~\cite{Stanev:1982au}.
The LPM effect is taken into account in \cor8 using a rejection sampling technique, where each interaction is rejected with a probability based on the difference between the interaction cross section with and without LPM suppression.
This approach is similar to the method used in \cor7~\cite{Heck:1998gr}, but it can now be used in air as well as in dense media.
The effect of the LPM suppression on \SI{100}{EeV} electromagnetic showers in air is shown in Figure~\ref{fig:long_100EeV}, based on a total of \num{5000} photon-induced showers simulated with and without the LPM effect in both \cor7 and \cor8.

\subsection{Thinning}
Thinning, introduced by Hillas~\cite{Hillas:1981}, is a technique that reduces the runtime of shower simulations, especially at the highest energies,
to a manageable timeframe. It is an example of a more general class of methods called \emph{Russian roulette} that has been used in other fields where radiation
transport is relevant~\cite{Garcia-Pareja:2021}. The general idea is to reduce the number of individually tracked particles in the shower by
\emph{pruning} the shower randomly, i.e., by discarding a large fraction of the secondaries at each vertex. The discarded particles are accounted for
by assigning statistical weights to the retained particles. This ensures that the expectation values of observables calculated from the pruned shower match those of the full shower. However, the cost of applying thinning is the introduction of \emph{artificial} statistical fluctuations in addition to the physical shower-to-shower fluctuations. It has been shown that these artificial fluctuations can be
reduced by introducing a weight limitation~\cite{Kobal:2001jx}, which prevents particles from obtaining weights larger than a prescribed value.

The implementation in \cor8 is as follows\footnote{Currently, thinning is only applied to electromagnetic interactions with a $1 \rightarrow 2$ splitting.}: two parameters determine the phase space of particles subject to thinning, the
\emph{threshold energy}, $\Eth$ (usually expressed as a fraction $\varepsilon = \Eth/E_{\mathrm CR}$ of the primary energy), and the \emph{maximum weight}, $\wmax$.
If the energy of the incoming particle, $E_0$,  is above the threshold, $E_0 > \Eth$, no thinning is applied, i.e., the secondaries are propagated further as usual. The early stage of the shower development, consisting of only a small fraction of the particles, is the dominant source
of shower-to-shower fluctuations. This stage largely determines the ``fate'' of the later development. In this regime, application of thinning would be detrimental. Secondaries with $E_0 \le \Eth$ are subject
to the thinning procedure. When weight limitation have not yet set in (to be defined later), one of the two secondaries is randomly chosen to be retained while
the other is discarded. The selection probability of the $i$-th secondary is defined to be proportional to its energy, where $E_1$ and $E_2$ denote the energies of the two secondary particles produced in the interaction,
\begin{equation}
    p_i = \frac{E_i}{E_1+E_2}. \label{eqn:hillas_probability}
\end{equation}
The weight of the retained particle is increased by a factor $f_i = 1/p_i$, i.e.
\begin{equation}
    w_i = w_0 f_i = \frac{w_0}{p_i},
\end{equation}
where $w_0$ denotes the weight of the incoming particle ($=1$ for particles that have not yet been subject to thinning yet).

Weight limitation is considered as follows: if, at any point, the potential new weight of any secondary, $w_i$, exceeds the maximum weight, $\wmax$, we resort
to \emph{statistical thinning}~\cite{Kobal:2001jx}, in which each secondary is considered for retention or discarding independently. In this setting, we have more freedom to alter the retention probabilities as desired, since the sum of the retention probabilities does not need to equal one. Here, we set
\begin{equation}
    p_i = \max\left(\frac{E_i}{E_1+E_2}, \frac{w_0}{\wmax}\right),
\end{equation}
so that
\begin{equation}
    w_i = \frac{w_0}{p_i} = \min\left(w_0 \frac{E_1+E_2}{E_i}, \wmax\right).
\end{equation}
As soon as the maximum weight is reached in a particular branch of the shower, all particles descending from that vertex are tracked again with the same weight, $\wmax$.

\begin{figure}
    \centering
    \includegraphics[width=\linewidth]{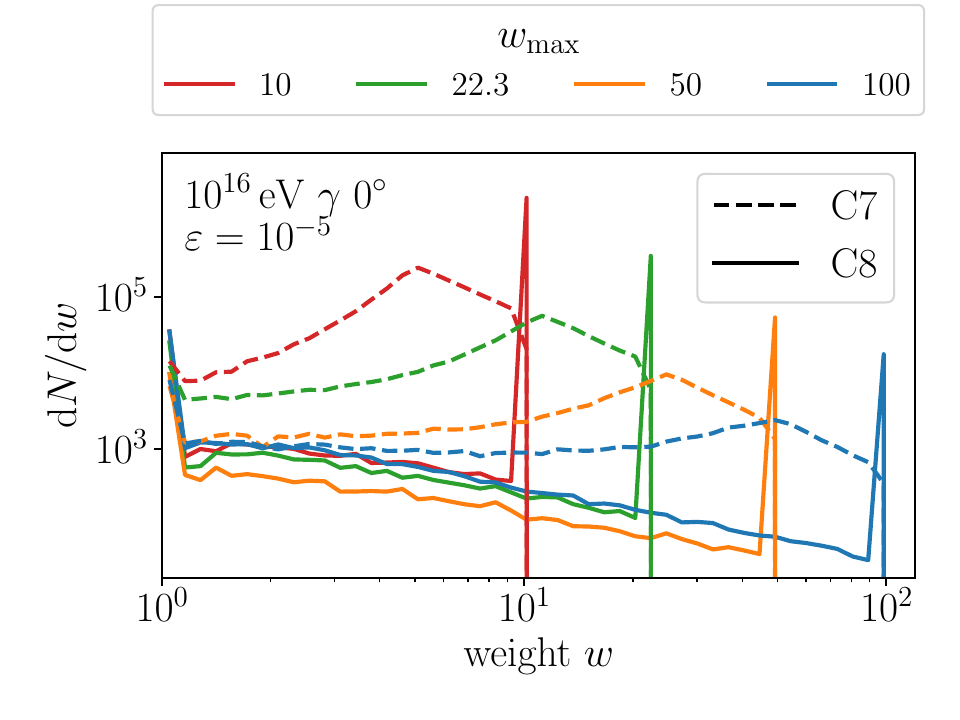}
    \caption{Weight distributions of photons due to thinning for different settings of the allowed maximum weight in comparison for \cor8 and \cor7.}
    \label{fig:thinning_wdist}
\end{figure}

\Cref{fig:thinning_wdist} shows a comparison of the weight distributions of photons obtained in $10^{16}\,\si{eV}$ photon showers using \cor8 and \cor7 with a thinning level $\varepsilon=10^{-5}$ and several maximum weight factors. In the high-weight range, \cor8
features a narrow peak, while \cor7 features a broad peak at a value of $\wmax / 2$. This difference is due to the different implementations of weight limitations. The current \cor8 implementation
switches to statistical thinning, which provides the freedom to adjust the retention probability to reach $\wmax$ exactly. \cor7, on the other hand, has no such freedom. The weight limitation effectively kicks
in earlier. When $p_i$ of \cref{eqn:hillas_probability} is too small to keep $w_i \leq \wmax$, no thinning is applied anymore, so that particles are ``stuck'' around $w \simeq \wmax/2$.

When comparing thinned and unthinned showers, it is important to note that enabling thinning alters the shower development. Even when using the same random number seed, a thinned shower will not be a pruned version of the shower obtained by disabling thinning. Since some of the particles and their interactions are skipped, the random numbers that would have been used there are used elsewhere, resulting in a different outcome.

To directly compare a full shower with one or more thinned realizations of the same shower,
we offer the option to enable thinning \emph{without discarding} particles (inspired by the multithinning method of \cor7~\cite{Heck:2014}). With this option, particles are assigned a weight of zero instead of being discarded, and are retained and further propagated. Thus, the random-number sequence remains undisturbed (the random numbers used for thinning are drawn from an independent sequence). By taking into account all particles with their weight $w \geq 1$, one can calculate observables of the thinned shower, as weight-zero particles do not contribute. Disregarding their weights and treating all particles as having a weight of one allows calculating observables of the full shower. Furthermore, different thinned realizations of the same shower can be obtained by running the simulation again with the same primary seed but a different one for the random number sequence used for thinning. A sufficiently large sample of these realizations is useful for studying the artificial fluctuations induced by thinning. For example, one could use it to test biases or sophisticated reconstruction algorithms that attempt to recover the real particle distribution given a thinned shower, as described in Refs.~\cite{Stokes:2011wf,Billoir:2008zz}.

\section{Hadronic showers}
\label{sec:hadronic_showers}

In a next step, we discuss and benchmark the simulation of hadronic showers in air with \cor8.

\subsection{Implementation}
In contrast to EM interactions, there is no closed theory for hadronic interactions. In other words, there is no unambiguous way to calculate predictions about particle production from first principles (QCD). In CORSIKA, phenomenological models are used to simulate hadronic showers. To understand the uncertainty resulting from imperfect modeling of hadronic interactions, it is necessary to support a variety of models. Currently, \cor8 provides interfaces to several widely used hadronic interaction models in cosmic-ray and air-shower physics, including EPOS-LHC~\cite{Pierog:2013ria}, QGSJet-II.04~\cite{Ostapchenko:2010vb}, SIBYLL~2.3d~\cite{Riehn:2019jet}, and, more recently, QGSJet-III~\cite{Ostapchenko:2024jsg}, EPOS-LHC-R~\cite{Pierog:2025ixr} (not yet available in the ``ICRC 2025 release''), and Pythia~8/Angantyr~\cite{Bierlich:2022pfr,CORSIKA:2025ops}\footnote{v8.315}. Additional hadronic interaction models may be included in future releases. Unlike models used in high-energy physics (HEP), which predominantly focus on high-$p_{\rm T}$ processes (e.g.\ Herwig++~\cite{Bahr:2008pv}), the models used for air shower calculations must be fully inclusive. To facilitate interaction with the HEP community, the interface between \cor8 and the hadronic models is designed to be generic. The two main interface routines \lstinline{getCrossSection} and \lstinline{doInteraction} only require the identification codes and four-momenta of the projectile and target particles in an arbitrary frame of reference. The transformation to the frame of reference of the respective model is done inside the interface in such a way that the overhead due to the transformations is minimal in the air shower context. Consequently, the same interface used in \cor8 can be used to provide predictions for accelerator experiments, facilitating the testing of the models against experimental data (Rivet~\cite{Bierlich:2019rhm}, MCplots~\cite{Karneyeu:2013aha}) without the need for additional code packages like CRMC~\cite{2021zndo...5270381U}.

The models listed above are high-energy (HE) hadronic interaction models specifically designed to describe interactions of hadrons in the multi-particle production regime ($\sqrt{s} \gtrsim \SI{10}{\GeV}$). At lower energies, particle production is dominated by discrete hadron resonances. Since the targets in air showers are nuclei, additional nuclear effects (e.g.\ Fermi motion of nucleons, intranuclear cascade, and fragmentation) become important when interaction energies reach the GeV scale and below. Not all HE models include these effects. Therefore, in \cor{} high- and low-energy (LE) interactions are distinguished. Currently, \cor8 provides an interface for FLUKA~\cite{refId0,Ferrari:2005zk} to handle LE hadronic interactions. The energy at which to switch between the HE and LE models varies among the different HE models and can be configured by the user. The default transition energy is \SI{79.4}{\GeV} for all supported HE models, except for Pythia~8/Angantyr for which a higher threshold of \SI{100}{\GeV} is used for technical reasons.

\subsection{Hadronic decays}
After their production, hadrons either interact again or decay. For the modeling of decays Pythia~8 is used in \cor8 and the lifetimes of hadrons are defined according to the Review of particle physics~2024~\cite{ParticleDataGroup:2024cfk,Rodrigues_Particle}. The probability whether a decay or an interaction occurs is determined by the interaction and decay lengths. In \cor7, which is strictly designed for cascades in air, the list of hadrons that interact before they can decay is short. With the exception of neutral pions, any hadron more short-lived than K$^0_{\rm short}$ is not tracked in \cor7 but forced to decay inside the generator. This may have an effect at energies beyond $100\,$EeV. In \cor8, which is designed to simulate cascades in any medium, in principle any hadron could interact and so all hadrons should be tracked and no generator-internal decays should be allowed. However, for the moment only interaction cross sections for long-lived hadrons are implemented and short-lived resonances are set to decay immediately after their production. This configuration was chosen because parameterizations of the interaction cross sections for short-lived hadrons are not easily available for all the interaction models. Since this configuration is very similar to \cor7, this choice also makes the comparison of shower observables much easier. A notable exception is QGSJet-II.04. For $\Lambda$ hyperons produced in showers simulated with this model, no interaction is implemented in \cor7, while in \cor8 the $\Lambda$s are treated as if they had been produced as neutrons.

\subsection{Particle history}\label{sec:cor8_lineage}
The uncertainty in modeling hadronic interactions, as mentioned above, has real consequences for cosmic-ray air shower experiments. It is an established fact that air shower simulations do not consistently describe the data of cosmic ray observatories~\cite{ArteagaVelazquez:2023fda,Albrecht:2021cxw}. The prevailing view is that this discrepancy stems from the incomplete modeling of HE hadronic interactions~\cite{Albrecht:2025kbb}. However, it is not yet clear which aspect of hadronic interactions is misrepresented. In fact, the exact mapping of hadronic interaction properties to air shower observables is not well understood. This is due to the immense increase in the number of degrees of freedom during the development of an air shower. In \cor7, all of these degrees of freedom are represented in the simulation, but they cannot be read out without significantly restructuring the code. \cor8 was designed specifically to lift this limitation. For the first time, it is now possible to inspect the entire history of every particle that reaches the ground~\cite{Alves:2021wiw,Reininghaus:2021zge}. This allows for much more detailed studies of the air shower development where the causal connection of each particle to the primary particle can be traced.

\subsection{Validation}

To validate the hadronic interaction framework implemented in \cor8, we compare key air-shower observables obtained with \cor8 to those obtained with the well-established \cor7 code. Since both frameworks provide interfaces to the same set of high- and low-energy hadronic interaction models, such a comparison allows us to isolate effects arising from the new simulation architecture and interfaces rather than from differences in the underlying physics models. Therefore, the focus of this validation is not to test agreement with experimental data; rather, it constitutes a consistency study between two independent implementations that nominally rely on the same underlying physics models.

For this purpose, we study vertical proton-induced air showers at a primary energy of \SI{e17}{eV}. We analyze the resulting showers, including the electromagnetic cascade generated by hadronic interactions, in terms of longitudinal and lateral observables for different particle components. High-energy hadronic interactions are modeled using EPOS-LHC, QGSJet-II.04, and SIBYLL-2.3d, while FLUKA is used as the LE interaction model. Identical thinning levels, energy cuts, and atmospheric conditions are applied in both simulation frameworks. We use a thinning level of \num{e-6}, with energy cuts of 10~MeV for electromagnetic particles and 300~MeV for hadrons and muons. Note that in \cor8, thinning is applied only to the electromagnetic component, whereas in \cor7 it is applied to the electromagnetic and hadronic components. Further information on the exact configuration is provided in \cref{sec:mc_config}. Statistical uncertainties are estimated using bootstrap resampling of the simulated shower sets.

\begin{figure}
    \centering
    \includegraphics[width=\linewidth]{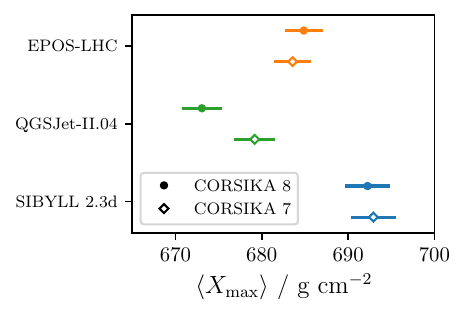}
    \caption{Average \Xmax for proton-induced showers at \SI{e17}{eV} simulated with \cor7 and \cor8 for different high-energy hadronic interaction models. Error bars indicate the statistical uncertainty of the mean.}
    \label{fig:Xmax_shift}
\end{figure}

First, we focus on the depth of the shower maximum, \Xmax. We compare the average \Xmax obtained from 1,000 \cor8 simulations with the result of 1,000 \cor7 simulations. The longitudinal shower profile, i.e., the energy deposited as a function of atmospheric depth, is used to determine \Xmax by fitting a parabola to the peak using the five neighboring bins on either side of the maximum bin. The results are shown in \cref{fig:Xmax_shift}. For EPOS-LHC and SIBYLL-2.3d, the results from both frameworks are consistent within statistical uncertainties. A small but significant deviation of \SI{\approx 6}{g\,cm^{-2}} is observed for QGSJet-II.04. This is likely related to the fact that for \cor7 the interaction of $\Lambda$-hyperons is not implemented.

\begin{table*}
\centering
\caption{Mean $\Xmax^p$ [g\,cm$^{-2}$] for different particle types and hadronic models. The quoted uncertainties are the uncertainties of the mean.}
\label{tab:mean_Xmax_particle_types}
\begin{tabular}{lccccc}
\toprule
Model & EAS Code & e$^\pm$ & $\mu^\pm$ & Hadrons & $\gamma$ \\
\midrule
SIBYLL-2.3d & 
\makecell{C7\\C8} &
\makecell{ 692 $\pm$ 2 \\  692 $\pm$ 2} &
\makecell{ 869 $\pm$ 2 \\  872 $\pm$ 2} &
\makecell{ 639 $\pm$ 2 \\ 644 $\pm$ 2} &
\makecell{ 725 $\pm$ 2 \\ 725 $\pm$ 2} \\
\\
QGSJet-II.04 & 
\makecell{C7\\C8} &
\makecell{677 $\pm$ 2 \\ 670 $\pm$ 2} &
\makecell{853 $\pm$ 2 \\ 845 $\pm$ 2} &
\makecell{622 $\pm$ 2 \\ 617 $\pm$ 2} &
\makecell{710 $\pm$ 2 \\ 703 $\pm$ 2} \\
\\
EPOS-LHC & 
\makecell{C7\\C8} &
\makecell{681 $\pm$ 2 \\ 682 $\pm$ 2} &
\makecell{853 $\pm$ 2 \\ 857 $\pm$ 2} &
\makecell{641 $\pm$ 2 \\ 647 $\pm$ 2} &
\makecell{714 $\pm$ 2 \\ 714 $\pm$ 2} \\
\bottomrule
\end{tabular}
\end{table*}

\begin{figure*}
    \centering
    \includegraphics[width=.85\linewidth]{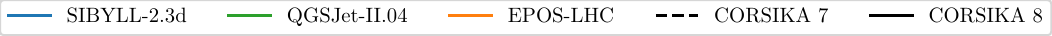}
    \includegraphics{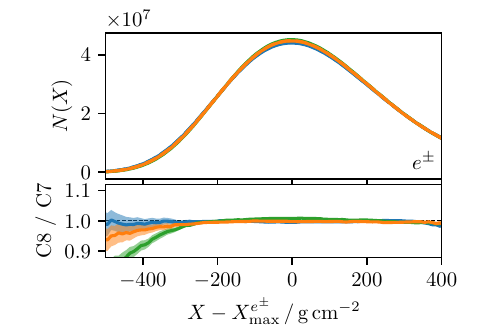}
    \includegraphics{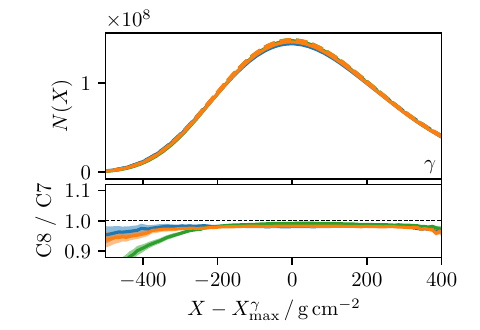}
    \includegraphics{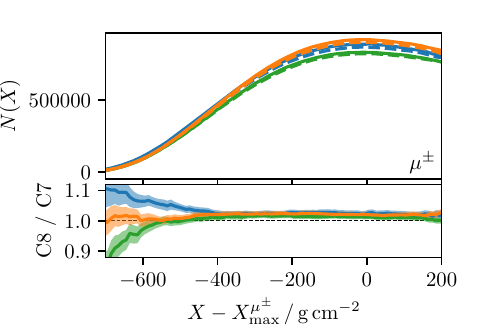}
    \includegraphics{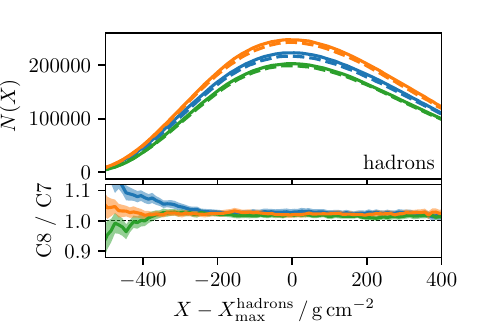}
    \caption{Median longitudinal profiles of 1,000 proton-induced vertical air showers with primary energy of \SI{e17}{eV} for various particle types (see indicator in plots) arising from hadronic cascades simulated with various HE hadronic interaction models and FLUKA as LE interaction model, both with \cor7 and \cor8. The shaded area shows the uncertainty of the median estimated via bootstrapping.}
    \label{fig:longprofiles_corr}
\end{figure*}

We now compare the longitudinal profiles of various particle types $p$, considering electrons and positrons, photons, muons, and hadrons separately. For each shower, the longitudinal profile is expressed relative to its maximum, i.e., as a function of $X - X_\mathrm{max}^p$. This allows for a direct comparison of the shower shapes, independent of a potential systematic shift between \cor7 and \cor8. The median longitudinal profiles of the different particle types, expressed in depth with respect to $\Xmax^p$, are shown in \cref{fig:longprofiles_corr}. Throughout most of the shower development, the ratio C8/C7 remains relatively flat, deviating by less than $\sim$\SI{5}{\percent}. Larger deviations primarily appear in the very early stages of the shower and near the end of the profile close to the ground level. These results demonstrate that the overall longitudinal evolution of the showers is very similar between \cor7 and \cor8. We also inspected the distribution of $X_\mathrm{max}^p$. A comparison of the values of the mean between the two codes is summarized in \cref{tab:mean_Xmax_particle_types}. The width of the $X_\mathrm{max}^p$ distributions is identical within the statistical uncertainties.

\begin{figure*}
    \centering
    \includegraphics[width=.85\linewidth]{figures/legend_box.pdf}
    \includegraphics{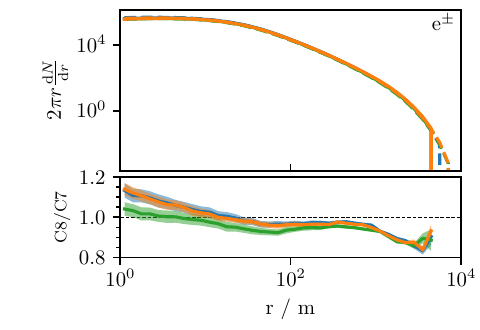}
    \includegraphics{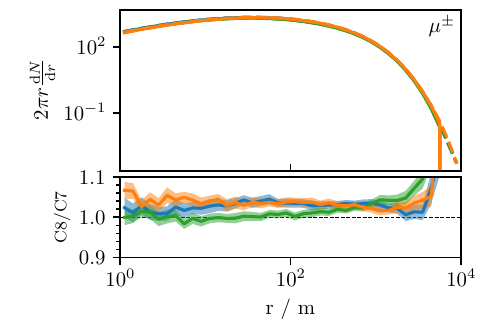}
    \includegraphics{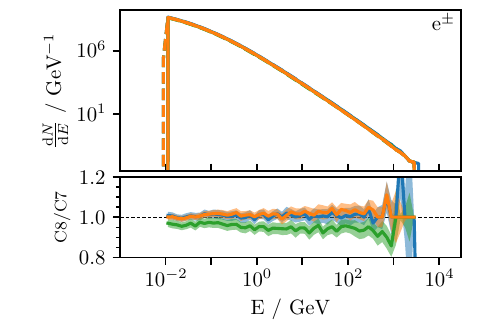}
    \includegraphics{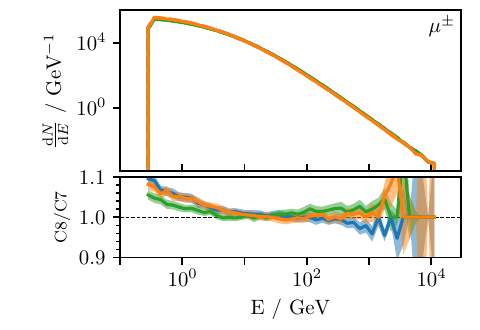}
    \caption{Top: Median lateral distributions of electrons and positrons (left) and muons (right) at ground level for the same showers as shown in \cref{fig:longprofiles_corr}. Bottom: Median energy spectra of electrons and positrons (left) and muons (right). The shaded areas show the uncertainty of the median estimated via bootstrapping.}
    \label{fig:ldfs}
\end{figure*}

Next, we examine shower observables at ground level, focusing on the lateral distribution of particles as a function of distance from the shower axis and their energy spectra. \Cref{fig:ldfs} shows the median lateral distributions and energy spectra for electrons and positrons, and muons from the same set of showers. Shaded areas indicate the statistical uncertainty of the median, estimated via bootstrap resampling. In general, the lateral distributions agree at the $\sim$\SI{10}{\percent} level for most particle species. The observed differences appear to depend on the HE interaction model. Although uncertainties at the 10\% level are non-negligible and warrant further investigation, we consider reaching agreement at this level between completely independent codes already a notable accomplishment.

\section{Cross-media showers}
The development of high-energy neutrino experiments in ice and water has created a need for a Monte Carlo code capable of simulating cross-media showers. One of the largest sources of background for these experiments are secondary high-energy muons generated by high-energy cosmic-ray-induced air showers~\cite{Barwick:2022vqt, KM3NeT:2009xxi}. When these muons lose a significant amount of energy, they tend to mimic the signal of an upward-going neutrino. Thus, simulations of showers from one medium to another are necessary for background estimation. Additionally, radio emission from cross-media showers have been observed in radio-based neutrino detectors in ice \cite{xwqy-yzrk}, constituting an important background and valuable calibration source, which requires detailed simulation.

Until now, most simulation programs, such as AIRES and \cor7, could only run particle showers in one medium. Geant4~\cite{GEANT4:2002zbu}, on the other hand, could run them in multiple media, but each of them only with constant density. Since none of the codes can handle both air and dense media simultaneously, most experiments opted to run the showers in two separate codes. For example, they combined Geant4 and \cor7 \cite{DeKockere:2024qmc}, saving particle data at the interface boundary and propagating it from one code to the other. While this approach has enabled detailed simulations of cross-media showers and their radio emission, a fully integrated solution would certainly be preferable.

\begin{figure}
    \centering
    \includegraphics[width=\linewidth]{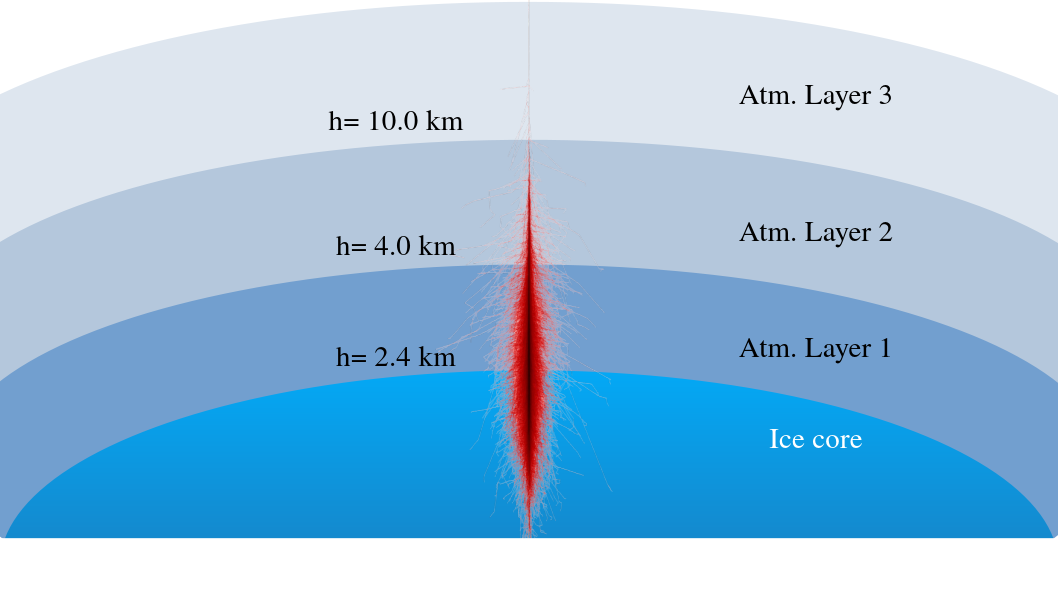}
    \caption{Simulation geometry for the cross-media shower example. The setup includes a five-layer exponential atmosphere and an ice volume starting at \SI{2.4}{km} above sea level. Only the first three atmospheric layers are shown.}
    \label{fig:air_ice_env}
\end{figure}

\cor8 solves this issue with its new environment setup. It is now possible to run showers in environments with multiple types of media of various shapes, each with different properties, all in one go. For testing purposes, we built a realistic environment found in Antarctica consisting of a five-layer atmosphere and an ice layer starting at \SI{2.4}{km} above sea level, as shown in Fig.~\ref{fig:air_ice_env}. To simulate a shower in this setup, the nuclear composition and density function, $\rho(h)$, must be provided, as well as conversions from grammage to length ($X(l, h_0)$) and from length to grammage ($l(X, h_0)$) for each medium. 

\begin{figure}
    \centering
    \includegraphics[width=0.99\columnwidth]{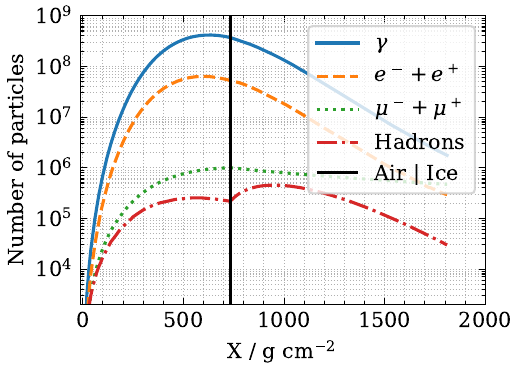}
    \caption{Longitudinal profile of a vertical \SI{100}{PeV} proton cross-media shower starting at the top of the atmosphere and intersecting an ice layer at an altitude of \SI{2.4}{km} above sea level, as shown by the black line.}
    \label{fig:cross_media_long}
\end{figure}

We simulated a vertical shower induced by a \SI{100}{\peta\eV} proton primary injected at the top of the atmosphere, with a magnetic field of (9.07, 0, 61.80) \SI{}{\micro\tesla} intersecting the air-ice interface at an altitude of \SI{2.4}{\km}. The shower was simulated using a thinning level of $10^{-6}$, a maximum weight of 50, an electromagnetic energy cut at \SI{0.2}{\MeV}, and a muon and hadronic energy cut at \SI{0.3}{\GeV}. The longitudinal profile of the different particles in the shower is shown in \cref{fig:cross_media_long}. While the profile from an electromagnetic shower would appear identical when plotted against grammage, even when transitioning from one medium to another, the hadronic component is sensitive to changes in density due to the interplay between the decay and interaction lengths. When the hadrons start propagating through ice, their interaction probability increases proportionally to the density increase (three orders of magnitude between air and ice). Consequently, hadrons start interacting with nuclei rather than decaying, producing the bump observed in \cref{fig:cross_media_long}. This also causes a slight increase in the electromagnetic component of the shower as $\pi^0$ particles decay into photons.

\begin{figure}
\centering
\includegraphics[width=0.99\columnwidth]{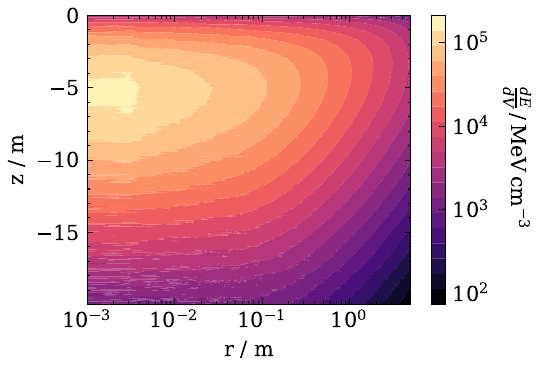}
\caption{Radial energy deposit of a 100 PeV proton vertical cross-media shower starting at the top of the atmosphere and intersecting an ice layer at an altitude of 2.4 km above sea level.}
\label{fig:cross_media_energy_deposit}
\end{figure}

Lastly, the energy deposition of the shower in ice is shown in \cref{fig:cross_media_energy_deposit}. The change in scale is evident, as the energy is primarily deposited in a cylinder measuring \SI{10}{\metre} in length and \SI{1}{\metre} in radius. This setup has also been used to test the results of \cor8 against those of \cor7 plus \textsc{Geant4} \cite{DeKockere:2022bto}. A good agreement is found between the frameworks \cite{CORSIKA:2023thf}. Work on implementing the calculation of radio emission from cross-media showers with \cor8 is currently ongoing.

\section{Electromagnetic emissions of showers}
The astroparticle physics community requires not only Monte Carlo simulations of particle cascades in air and dense media, but also high-fidelity simulations of the electromagnetic emissions of these showers.

\subsection{Cherenkov light emission}
For several decades, the detection of extensive air showers via optical light, particularly Cherenkov and fluorescence light emission, has been a cornerstone of high-energy cosmic-ray and gamma-ray astronomy. Advances in photon detection technologies, optical systems, and atmospheric monitoring have enabled increasingly precise air shower measurements using imaging Cherenkov and fluorescence light telescopes. These developments require simulation tools that can accurately and flexibly model the production and propagation of optical light in realistic experimental environments.
Earlier generations of air-shower simulation codes tightly coupled optical light production to the shower simulation itself, limiting extensibility and making it difficult to adapt the implementation to novel detector concepts or complex geometries. Therefore, in \cor8, the simulation of optical light is being redesigned as a fully modular, first-class component of the framework, following the same design principles as other observational processes.

Simulating optical light can account for a significant portion of the total runtime. To address this issue, the optical light module currently being developed in \cor8 provides dedicated interfaces that allow for the early rejection of particle tracks and optical photons that do not contribute to measurable signals. This feature is especially important  for high-energy cascades and for simulating of edge cases, in which only a portion of the photon field is within the sensitive area.

These interfaces allow the emission and propagation components to apply experiment-specific criteria, such as geometric visibility, wavelength acceptance, and detector thresholds, before incurring the full computational cost of photon generation and transport. Therefore, particle tracks that cannot produce Cherenkov light under the local refractive conditions or whose fluorescence emission cannot reach any active detector element can be discarded early on. Similarly, optical photons can be removed during propagation once it is determined that they no longer contribute to the detector response.

This approach maintains the purely observational nature of the optical processes while significantly reducing runtime for realistic detector geometries. It enables the efficient, large-scale production of simulated events and systematic studies without compromising physical accuracy in the measurable phase space.

From a computational perspective, a new approach is taken that enables easier compiler-based vectorization by disconnecting photons from the main simulation loop. This approach also lends itself to an asynchronous processing model for photons. For example, a GPU-based implementation is available~\cite{Reininghaus:2021qoa}.

The optical light module currently exists as a fork of the \cor8 project and has been validated against the established implementations used in \cor7 and in experiment-specific simulation chains. Benchmarks demonstrate good agreement in photon yields, ground distributions, and longitudinal profiles for both Cherenkov and fluorescence light. This confirms the physical consistency of the new implementation. The decoupled design also substantially improves maintainability and extensibility, providing a solid foundation for future developments~\cite{CORSIKA:2023sfd}. The upcoming Cherenkov Telescope Array Observatory (CTAO)~\cite{Gueta:2021vrf}, in particular, will require simulations with a fidelity that is not currently achievable with \cor7, as \cor7 is limited by the concept of five-layer exponential atmospheres. Similarly, legacy codes do not currently support Cherenkov light simulations from upward or atmosphere-skimming air showers.

Work is underway to fully integrate the optical light module in the mainline branch of the project.

\subsection{Radio emission}
Technological advancements in digital signal processing, combined with the excellent understanding of the radio emission phenomena have led to a significant progress in radio detection techniques for air showers over the past two decades \cite{Huege:2016veh}. As a result, experiments have been proposed to study the radio signals emitted from ``non-standard'' air showers. Examples include air showers crossing media from air to ice, refracting and/or reflecting radio signals at a boundary crossings, and experiments with an ever-increasing number of antennas. The community's leading standards for the simulation of the radio emission are the \cor7 code with its CoREAS extension \cite{Huege:2013vt} and the ZHAireS code \cite{Alvarez-Muniz:2011ref}. However, both of these solutions suffer from the flexibility limitations of the underlying \cor7 and AIRES simulation codes.

To address these limitations, a radio module has been designed and developed as an integral part of \cor8. Following the modular design principles of the framework, the radio emission simulation is decomposed into four independent, user-configurable components: optional filtering of particle tracks, a formalism for calculating the  radio emission, propagation of radio signals through complex media (potentially with multiple paths), and signal reception by antennas. Currently, the ZHS \cite{Zas:1991jv, Alvarez-Muniz:2010wjm} and CoREAS (``endpoints'') \cite{James:2010vm} algorithms in the time domain along with propagators that use a straight-ray approximation and time domain ``ideal antennas'', are fully implemented. One of the main advantages of the module's architecture is its ability to include custom-made propagators designed for specific experimental setups. For example, it can handle air–ice transitions in cross-media showers, as discussed in the previous section and demonstrated in Ref.~\cite{CORSIKA:2023thf, Coleman:2024cln}. The radio process is implemented as a purely observational component and therefore does not influence the development of the air shower.

The radio module has been extensively validated as both a stand-alone feature and a part of \cor8. A comprehensive description of the module design and implemented algorithms, as well as detailed validation and benchmark studies, can be found in Ref.~\cite{Alameddine:2024cyd}.
When sufficiently small step lengths are enforced, the total radio emission from \cor8 agrees with \cor7 to better than \SI{10}{\%} in the \SIrange{30}{80}{MHz} band and is negligible in the \SIrange{50}{350}{MHz} band.
Therefore, we conclude that the radio module within \cor8 is ready for use in physics applications.

While the standard \textit{radio process} is sufficient for most air shower applications, \cor8 is currently being extended to also handle radio emission and propagation in complex media. For smoothly-inhomogeneous environments in which media properties change on length scales that are large compared to the wavelength, geometric optics is an adequate description and radiation can be thought of as propagating along curved rays. \cor8 is being equipped with a general numerical raytracing propagator that can be used as part of the \textit{radio process}~\cite{CORSIKA:2025rpb}. This approach will allow for the accurate simulation of radio signals from neutrino-induced in-ice showers. To handle even more complicated situations where inhomogeneities occur on the scale of the wavelength, \cor8 is being interfaced with the external \textsc{Eisvogel} package~\cite{Windischhofer:2023Jn, Windischhofer:2024Ba}. \textsc{Eisvogel} provides a full-electrodynamics treatment of radio emission and propagation based on Green's functions \cite{CORSIKA:2025rpb}. This allows for accurate simulations in complex geometries and highly inhomogeneous media, such as radiation propagating in shallow ice, where wave-optics phenomena become relevant. Recently, \cor8 and \textsc{Eisvogel} have been used to model radio signals from air shower cores impacting polar ice sheets, enabling direct comparisons to experimental observations of in-ice Askaryan radiation~\cite{xwqy-yzrk}.

\section{Performance benchmarks}
We benchmarked not only the simulation predictions of \cor8, but also its its runtime performance against \cor7. Here\footnote{For these benchmarks, a slightly optimized version of the ICRC2025 release was used.}, we showcase its performance on a dedicated cluster node running AlmaLinux~9.6. The processor was a 12-core Intel Xeon running at $2.40\,$GHz. Only AVX vector extensions were available. Both codes were compiled with GCC 11.5 and the following optimization flags: \lstinline{-O3 -mavx -march=native -mtune=native}, which ensured exploitation of the available vector instruction set and architecture-specific optimizations.
To minimize hardware-induced variability in runtime measurements, Intel Turbo Boost and hyperthreading were disabled on the cluster node. Similarly, I/O operations were performed on a local disk to minimize latency.

To reduce statistical uncertainty on the measured runtimes, 10 runs were performed for each benchmark configuration. Each run consisted of 10, or 100 showers for short runtimes, to minimize fixed overheads, such as initialization, while keeping the overall computational cost reasonable. To get a more representative performance measurement, a distinct random number generator seed was assigned to each shower, so that showers in different runs but in the same position in the sequence start with the same seeds. Our benchmarks focus primarily on pure hadronic showers, with all electromagnetic particles removed from the simulation. A brief qualitative comment on electromagnetic shower performance and associated challenges is provided in a separate subsection. Identical physics configurations and primary showers were used in both \cor7 and \cor8. All benchmarks correspond to vertical showers.

\subsection{Hadronic showers}
 
 \begin{figure}
    \centering
    \includegraphics[width=\linewidth]{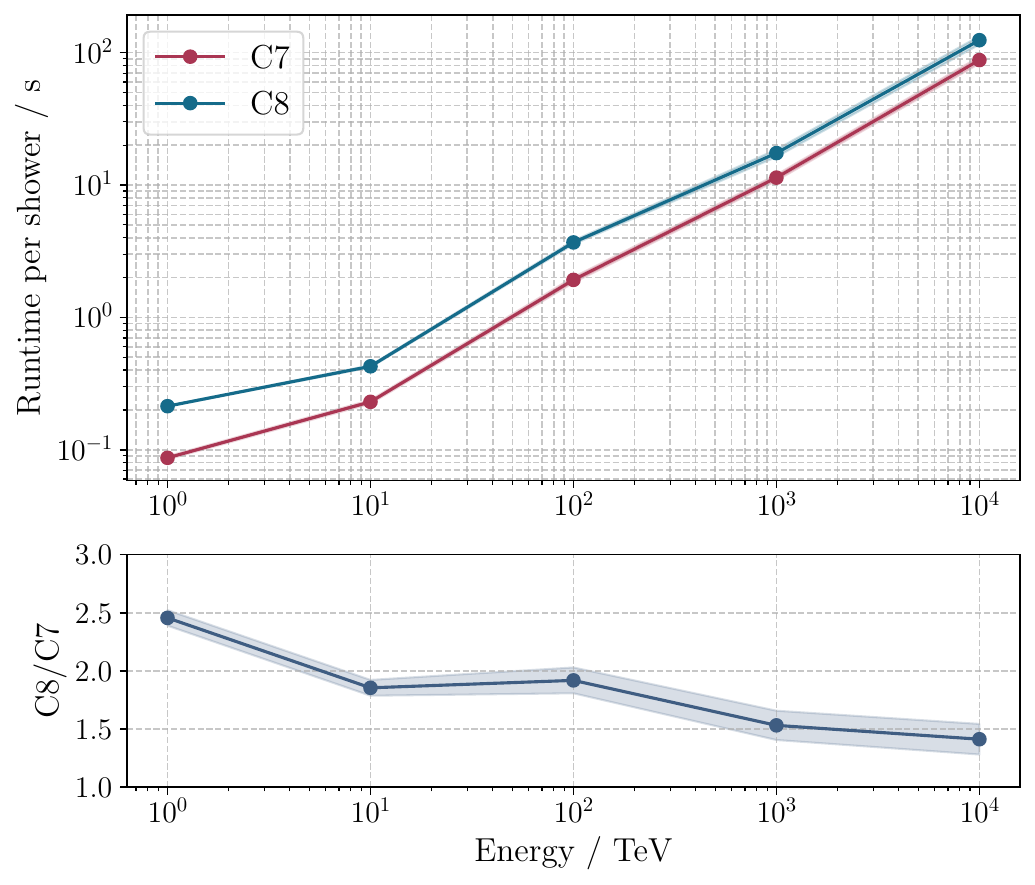}
    \caption{Runtime per shower for proton-induced hadronic showers from $1\,$TeV to $10\,$PeV with all electromagnetic particles cut. Each point represents the mean runtime per shower, computed from 10 runs of either 10 or 100 showers. The shaded bands indicate the standard deviation over 10 runs.}
    \label{fig:bench_TeV}
\end{figure}

Figure \ref{fig:bench_TeV} shows the runtime per shower for proton-induced hadronic showers ranging from $1\,$TeV to $10\,$PeV using the SIBYLL~2.3d model for high-energy hadronic interactions. In this energy range, \cor8 is between 1.5 and 2.5 times slower than \cor7. For shorter showers (lasting only a few seconds), the ratio is closer to 2.5, likely due to larger initialization overheads in \cor8. For more energetic showers, the ratio decreases to about 1.5. 

\begin{figure}
    \centering
    \includegraphics[width=\linewidth]{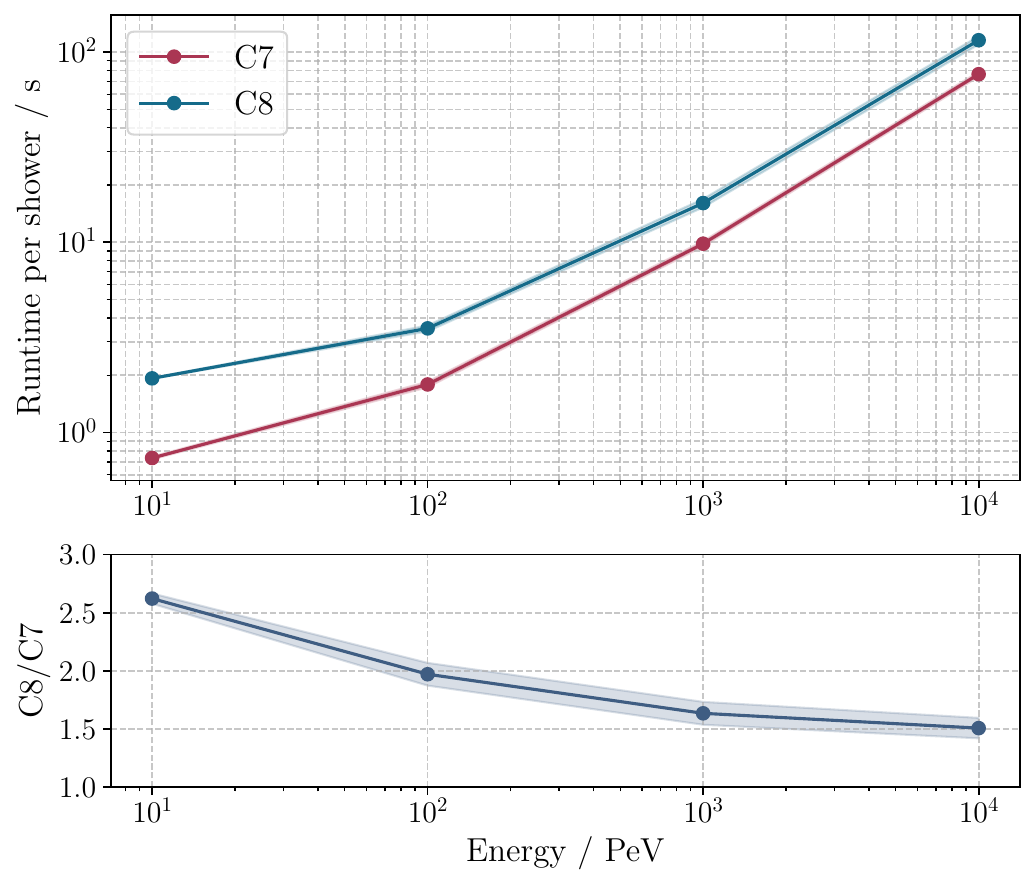}
    \caption{Same as \cref{fig:bench_TeV}, but for energies ranging from $10\,$PeV to $10\,$EeV. An additional cut was added for hadronic particles below $1\,$TeV to reduce the overall runtimes for \cor7 and \cor8.}
    \label{fig:bench_PeV}
\end{figure}

The results for higher-energy showers, ranging from $10\,$PeV to $10\,$EeV, are presented in \cref{fig:bench_PeV}. An additional cut on hadrons with energies below $1\,$TeV was applied here to keep runtimes at a manageable level. The same qualitative behavior is observed, with comparable runtime ratios between \cor7 and \cor8, and a similar dependence on the primary energy.

It is expected that \cor8 is somewhat slower than \cor7, as \cor7 has already undergone decades worth of optimization, whereas \cor8 has not yet been systematically optimized. Therefore, the results shown here only constitute an initial performance baseline. Furthermore, it should be kept in mind that \cor7 is heavily hand-optimized for a limited number of particular use cases. This makes it very performant but also limits its flexibility. Thus, it is expected that the fully flexible approach of \cor8 cannot entirely reach the performance of \cor7. We consider a factor of 1.5 to be very acceptable, demonstrating that the chosen code design is fully adequate.

The benchmarks were repeated using QGSJet-II.04 and EPOS-LHC as hadronic interaction models. For QGSJet-II.04, runtime ratios similar to those obtained with SIBYLL~2.3d were observed. However, for EPOS-LHC, \cor8 was approximately 3.5 times slower than \cor7, a difference that will be investigated further in the future.

\subsection{Electromagnetic showers}

While hadronic showers in \cor8 perform only slightly worse than in \cor7, electromagnetic showers currently experience a roughly one order of magnitude slowdown in \cor8 compared to \cor7. The main difference between the approaches of \cor8 and \cor7 for simulating the electromagnetic cascade is that \cor8 uses PROPOSAL, which is interfaced by the \cor8 core functionality in a generic way. In contrast, \cor7 uses a strongly integrated custom version of EGS4 that handles a much larger part of the simulation than PROPOSAL does in \cor8. We are investigating the exact source of the slowdown, which could reside in the generic interface between \cor8 and PROPOSAL, the performance of PROPOSAL, or the tracking of electromagnetic particles in magnetic fields. As these investigations are still ongoing, no dedicated electromagnetic benchmark results are presented here.

\section{Discussion}
We have successfully validated \cor8 against \cor7 for both electromagnetic and hadronic air showers. As shown in the preceding sections, the agreement between the two codes is generally within $\sim$\SI{10}{\percent}. These validation studies demonstrate that the core physics implementations of \cor8 reproduce the well-established results of \cor7 for standard air shower scenarios and provide a solid baseline for further developments. For this purpose, \cor8 includes a standard application called \texttt{c8\_air\_shower} for air-shower simulations that closely mimics the traditional \cor7 setup. This application was used for the comparisons presented in this work.

At the same time, \cor8 is still subject to performance limitations and further optimization is ongoing. For this reason, the current recommendation is to continue using \cor7 for standard air shower simulations in the atmosphere of the Earth. However, for applications that are difficult or impractical to address within the constraints of the \cor7 code base, \cor8 is now the recommended choice. New investments in the legacy code base of \cor7 might not yield a favorable return in the long run.

 To demonstrate the expanded applicability of \cor8, we present example applications involving particle cascades in media other than air, such as in the Martian atmosphere and water. These examples demonstrate the flexibility of the \cor8 framework with respect to atmospheric profiles, material properties, and geometry, and as its ability to handle cross-media showers in a unified and consistent way. Additionally, \cor8 provides the foundation for detailed studies of radio signal generation and propagation in dense media, such as ice, which is particularly relevant for neutrino detection experiments. An example of radio propagation in ice based on \cor8 can be found in Ref.~\cite{Coleman:2024cln}.

At the current stage of development, some physical aspects are treated approximately. In particular, muon decays are modeled without polarization, which alters the angular distribution of decay electrons and positrons~\cite{Michel:1949qe}. This induces a small effect on the resulting e$^\pm$ energy spectra, cf.\ Ref.~\cite{Lipari:1993hd}. However, we emphasize that similar limitations exist in \cor7. The treatment of the electromagnetic cascade in \cor7 relies on EGS4, subsequent developments and refinements, such as EGS5~\cite{Hirayama:2005} and EGSnrc~\cite{EGSnrc}, were never backported.

\section{Conclusions and Outlook}
We have presented \cor8, a modern and modular framework for simulating extensive air showers and particle cascades in arbitrary media. Motivated by the long-term sustainability challenges and architectural limitations of \cor7, \cor8 was developed as a complete redesign rather than an incremental update. Its modular architecture has well-defined interfaces between physics processes, environment models, and geometry descriptions. This enables maintainable, extensible, and flexible simulations across a wide range of use cases.

The implemented electromagnetic and hadronic shower simulations have been validated through detailed comparisons with \cor7. For a broad set of observables, including longitudinal and lateral particle distributions and energy spectra, \cor8 reproduces the established results of its predecessor to a degree better than \SI{10}{\percent}. This is a significant achievement but also warrants further investigation.

Beyond reproducing traditional air-shower simulations, \cor8 is designed to support a broader range of applications. Its flexible environment model enables simulations in heterogeneous and non-standard media as well as arbitrarily complex geometries. Its modular process structure facilitates the integration of new interaction models. Modern data formats and accompanying Python analysis tools simplify integration into contemporary analysis pipelines.

Future work will focus on expanding the range of supported hadronic interaction models and enhancing the performance of the electromagnetic cascade simulation in particular. Thanks to its modular design and modern infrastructure, \cor8 provides a robust and future-proof foundation for air shower and cascade simulations in current and next-generation astroparticle physics experiments.

\appendix

\section{Simulation setup}
\label{sec:mc_config}
The \cor8 simulations are performed using the \texttt{c8\_air\_shower} standard application, which closely replicates the traditional \cor7 setup. The simulation uses a five-layer atmosphere based on Keilhauer's parametrization of the US standard atmosphere, as well as north-pointing  magnetic field of \SI{50}{\micro\tesla}. All particles are propagated until they reach the observer level at sea level.

Unless stated otherwise, a thinning level of \num{e-6} is adopted, with a maximum allowed weight set to half of the optimal value (Kobal's optimum) to match the typical behavior of \cor7. Depending on the simulation setup, particle energy cuts vary and are specified below. In particle tracking, a maximum deflection of 0.2 radians is permitted.

For electromagnetic showers with a primary energy of \SI{100}{TeV}, the energy cuts are 1~MeV for electromagnetic particles, 500~MeV for hadrons and muons, and 300~MeV for tau leptons. A typical command to run \cor8 simulations is:
\begin{lstlisting}[language=bash,breaklines=true]
c8_air_shower --pdg 22 --energy 1e5 --emcut 0.001
    --hadcut 0.5 --mucut 0.5 --nevent 1
    --seed {seed} --filename {filename}
    --observation-level 0e3
\end{lstlisting}

For electromagnetic showers with a primary energy of \SI{100}{EeV}, the energy cuts are 100~GeV for electromagnetic particles, 100~GeV for hadrons and muons, and 300~MeV for tau leptons. A typical command is:
\begin{lstlisting}[language=bash,breaklines=true]
c8_air_shower --pdg 22 --energy 1e11
    --emcut 100e3 --hadcut 100e3 --mucut 100e3
    --nevent 1 --seed {seed}
    --filename {filename} --observation-level 0e3
\end{lstlisting}

For hadronic showers, the energy cuts are 10~MeV for electromagnetic particles, and 300~MeV for hadrons, muons, and tau leptons. A typical command is:
\begin{lstlisting}[language=bash,breaklines=true]
c8_air_shower -p 2212 -E 1e8 -z 0 -N 1
    --emcut 0.01 --mucut 0.3 --hadcut 0.3
    --max-weight 0 -s {seed} -M SIBYLL-2.3d
    -f {filename}
    --disable-interaction-histograms
\end{lstlisting}

\cor7 input cards for the electromagnetic showers have the following form:
\begin{lstlisting}[language=bash,breaklines=true]
RUNNR   {run_idx}
EVTNR   {first_event_idx}
NSHOW   {n_show}
PRMPAR  {primary}
ERANGE  100.E9  100.E9
THETAP  0.     0.
PHIP    -180.  180.
SEED    {seed_1}   0   0
SEED    {seed_2}   0   0
OBSLEV  0.E0
MAGNET  50  0
ELMFLG  F   T
ECUTS   100.E3 100.E3 100.E3  100.E3
LONGI   T  10.  F  T
ATMOD   17
EXIT
\end{lstlisting}

and for the hadronic showers the following form:
\begin{lstlisting}[language=bash,breaklines=true]
RUNNR   {run_idx}
EVTNR   {first_event_idx}
NSHOW   1
PRMPAR  14
ERANGE  1.0E8  1.0E8
THETAP  0.  0.
PHIP    0.  0.
SEED    {seed_1}   0   0
SEED    {seed_2}   0   0
OBSLEV  0
MAGNET  50 0
ECUTS   0.3 0.3 0.01  0.01
ELMFLG  F   T
LONGI   T  10.  T  T
THIN    1.0e-06 100. 0.
THINH   1.00 1.00
ATMOD   17
EXIT
\end{lstlisting}

\section{Dependencies}

For proper attribution and reproducibility, we list the main software libraries used by \cor8 and in the analysis presented in this work.

\subsection{\cor8}
\begin{itemize}
    \item \texttt{Apache Arrow}~\cite{arrow}
    \item \texttt{Boost::histogram}~\cite{Dembinski:2020dic,Schreiner:2020,boost_histogram}
    \item \texttt{CLI11}~\cite{CLI11}
    \item \texttt{cnpy}~\cite{cnpy}
    \item \texttt{Eigen}~\cite{Eigen}
    \item \texttt{particle} (v0.25.1)~\cite{Rodrigues_Particle}
    \item \texttt{spdlog}~\cite{spdlog}
    \item \texttt{yaml-cpp}~\cite{yaml_cpp}
\end{itemize}
\subsection{Software used for this paper}
\begin{itemize}
    \item \texttt{boost-histogram} (python)
    \item \texttt{matplotlib}~\cite{Hunter:2007ouj,matplotlib_zenodo}
    \item \texttt{mplhep}~\cite{mplhep_zenodo}
    \item \texttt{numpy}~\cite{Harris:2020xlr}
    \item \texttt{pandas}~\cite{pandas}
\end{itemize}

\sloppy
\section*{Acknowledgements}
\noindent
We thank T.~Sjöstrand, L.~Lönnblad, and the Pythia 8 collaborators for their support in the implementation of Pythia 8/Angantyr in CORSIKA 8.

This research was funded by the Deutsche Forschungsgemeinschaft (DFG, German Research Foundation) – Projektnummer 445154105 and Collaborative Research Center SFB1491 ``Cosmic Interacting Matters - From Source to Signal''. This research has been partially funded by the German Federal Ministry of Research, Technology and Space (BMFTR) and the state of North Rhine-Westphalia through the Lamarr Institute for Machine Learning and Artificial Intelligence. We acknowledge support through project UNAM-PAPIIT IN114924.

This work has also received financial support from Ministerio de Ciencia e Innovación/Agencia Estatal de Investigación (PRE2020-092276). A.~Coleman is supported by the Swedish Research Council (Vetenskapsrådet) under project no.~2021-05449. C.~Glaser is supported by the Swedish Research Council (Vetenskapsrådet) under project no.~2021-05449, and the European Union (ERC, NuRadioOpt, 101116890). D.~Baack is supported by the European Union (ERC, NuRadioOpt, 101116890). F.~Riehn received funding from the European Union’s Horizon 2020 research and innovation programme under the Marie Skłodowska-Curie grant agreement No.~101065027. P.~Windischhofer and C.~Deaconu thank the NSF for Award 2411662. R.~Privara’s work on this paper was financially supported by the Ministry of Education of the Czech Republic and the project MEYS Infra Auger LM2023032. J.~Lazar is supported by the Belgian Fonds de la Recherche Scientifique (FRS-FNRS). The authors acknowledge support by the High Performance and Cloud Computing Group at the Zentrum für Datenverarbeitung of the University of Tübingen, the state of Baden-Württemberg through bwHPC and the DFG through grant no.~INST 37/935-1 FUGG. The computations were partially carried out on the PLEIADES cluster at the University of Wuppertal, which was supported by the DFG (grant no.~INST 218/78-1 FUGG) and the BMFTR.

\section*{Author contributions (CRediT)}

\textbf{J.M.~Alameddine}: Conceptualization, Methodology, Software, Validation, Formal analysis, Investigation, Writing - Original Draft, Visualization,
\textbf{J.~Albrecht}: Investigation, Resources, Supervision, Funding acquisition,
\textbf{A.A.~Alves Jr.}: Conceptualization, Methodology, Software, Formal analysis, Supervision, Project administration,
\textbf{J.~Ammerman-Yebra}: Conceptualization, Methodology, Software, Validation, Writing - Original Draft,
\textbf{L.~Arrabito}: Supervision,
\textbf{D.~Baack}: Conceptualization, Methodology, Software, Validation, Investigation, Resources, Project administration,
\textbf{A.~Coleman}: Conceptualization, Methodology, Software, Validation,
\textbf{C.~Deaconu}: Resources, Funding acquisition,
\textbf{H.~Dembinski}: Conceptualization, Methodology, Software, Investigation, Supervision,
\textbf{D.~Elsässer}: Supervision,
\textbf{R.~Engel}: Conceptualization, Supervision, Funding acquisition,
\textbf{A.~Faure}: Software, Formal analysis, Writing - Original Draft, Visualization,
\textbf{C.~Gaudu}: Conceptualization, Methodology, Software, Validation, Formal analysis, Investigation, Data curation, Writing - Review \& Editing, Visualization,
\textbf{C.~Glaser}: Conceptualization, Resources, Writing - Review \& Editing, Supervision, Project administration, Funding acquisition,
\textbf{M.~Gottowik}: Software, Validation, Formal analysis, Investigation, Writing - Original Draft, Visualization,
\textbf{T.~Huege}: Conceptualization, Methodology, Validation, Investigation, Resources, Writing - Original Draft, Writing - Review \& Editing, Supervision, Project administration, Funding acquisition,
\textbf{K.H.~Kampert}: Conceptualization, Validation, Resources, Writing - Review \& Editing, Supervision, Project administration, Funding acquisition,
\textbf{N.~Karastathis}: Conceptualization, Methodology, Software, Validation, Formal analysis, Investigation, Writing - Original Draft, Visualization,
\textbf{J.~Lazar}: Software, Validation, Writing - Original Draft,
\textbf{L.~Nellen}: Validation, Resources, Project administration,
\textbf{D.~Parello}: Supervision,
\textbf{T.~Pierog}: Conceptualization, Methodology, Writing - Review \& Editing, Supervision, Project administration,
\textbf{R.~Privara}: Software, Validation, Investigation,
\textbf{M.~Reininghaus}: Conceptualization, Methodology, Software, Validation, Formal analysis, Investigation, Writing - Original Draft, Writing - Review \& Editing, Visualization, Supervision, Project administration,
\textbf{W.~Rhode}: Investigation, Resources, Supervision, Funding acquisition,
\textbf{Riehn}: Conceptualization, Methodology, Software, Validation, Formal analysis, Investigation, Writing - Original Draft, Writing - Review \& Editing, Visualization, Supervision,
\textbf{M.~Sackel}: Conceptualization, Software, Validation,
\textbf{P.~Sampathkumar}: Validation, Investigation,
\textbf{A.~Sandrock}: Conceptualization, Methodology, Software, Validation, Formal analysis, Investigation, Writing - Original Draft, Writing - Review \& Editing, Visualization, Supervision
Schmidt: Software,
\textbf{J.~Soedingrekso}: Conceptualization, Methodology, Software, Validation, Formal analysis, Investigation,
\textbf{R.~Ulrich}: Conceptualization, Methodology, Software, Supervision, Project administration,
\textbf{P.~Windischhofer}: Software, Validation, Investigation, Writing - Review \& Editing,
\textbf{B.~Yue}: Software, Validation, Investigation.

\bibliographystyle{elsarticle-num}
\bibliography{literature.bib}

\end{document}